\documentclass[usenatbib,referee]{mn2e}
\usepackage{epsf}
\usepackage{graphicx}

\def\eps@scaling{1.0}%
\newcommand\epsscale[1]{\gdef\eps@scaling{#1}}%
\newcommand\plotone[1]{%
 \centering
 \leavevmode
 \includegraphics[width={\eps@scaling\columnwidth}]{#1}%
}%

\def\mnras{MNRAS}
\def\apjs{ApJS}

\def\beq{\begin{equation}}
\def\eeq{\end{equation}}
\def\bey{\begin{eqnarray}}
\def\eey{\end{eqnarray}}

\def\mpc{\, h^{-1}{\rm {Mpc}}}

\def\mpci{\, h{\rm {Mpc}}^{-1}}
\def\kpc{\, h^{-1}{\rm {kpc}}}

\def\kms{\,{\rm {km\, s^{-1}}}}

\def\msun{\, h^{-1}{\rm M_\odot}}

\def\gs{\mathrel{\raise1.16pt\hbox{$>$}\kern-7.0pt
\lower3.06pt\hbox{{$\scriptstyle \sim$}}}}
\def\ls{\mathrel{\raise1.16pt\hbox{$<$}\kern-7.0pt
\lower3.06pt\hbox{{$\scriptstyle \sim$}}}}
\def\gtsima{$\; \buildrel > \over \sim \;$}
\def\ltsima{$\; \buildrel < \over \sim \;$}
\def\prosima{$\; \buildrel \propto \over \sim \;$}
\def\gsim{\lower.5ex\hbox{\gtsima}}
\def\lsim{\lower.5ex\hbox{\ltsima}}
\def\simgt{\lower.5ex\hbox{\gtsima}}
\def\simlt{\lower.5ex\hbox{\ltsima}}
\def\simpr{\lower.5ex\hbox{\prosima}}
\def\la{\lsim}
\def\ga{\gsim}

\title
[Reconstructing the cosmic density field] {Reconstructing the
cosmic density field with the distribution of dark matter halos}
\author[Huiyuan Wang et al.]
   {\parbox[t]{\textwidth}{
       Huiyuan Wang$^{1,2,4}$\thanks{E-mail: whywang@mail.ustc.edu.cn},
       H.J. Mo$^{1}$,
       Y.P. Jing$^{3,4}$,
       Yicheng Guo$^{1}$,
       Frank C. van den Bosch$^{5}$
       and
       Xiaohu Yang$^{3,4}$
}\\
            $^1$Department of Astronomy, University of Massachusetts,
            Amherst MA 01003-9305, USA\\
            $^2$Center for Astrophysics, University of Science and
                Technology of China, Hefei, Anhui, China\\
            $^3$Shanghai Astronomical Observatory; the Partner Group of MPA,
            Nandan Road 80, Shanghai 200030, China \\
            $^4$Joint Institute for Galaxy and Cosmology (JOINGC) of SHAO
            and USTC\\
            $^{5}$Max-Planck-Institute for Astronomy, K$\ddot{o}$nigstuhl 17, D-69117 Heidelberg, Germany
}

\date{
Accepted ........
Received .......;
in original form ......}

\pubyear{2008}

\begin{document}
\maketitle \label{firstpage}

\begin{abstract}
  We develop a new method to reconstruct the cosmic density field from
  the   distribution of  dark  matter  haloes   above a  certain  mass
  threshold.  Our motivation  is that  well-defined samples of  galaxy
  groups/clusters,   which  can be  used  to represent   the dark halo
  population, can now  be   selected from large  redshift  surveys  of
  galaxies, and our ultimate goal  is to use  such data to reconstruct
  the cosmic density field  in the local universe.  Our reconstruction
  method starts with a sample of dark matter haloes above a given mass
  threshold.   Each volume element in  space is assigned to the domain
  of the nearest halo  according to a distance  measure that is scaled
  by the virial  radius of the halo.   The distribution of the mass in
  and around dark matter haloes of a given mass  is modelled using the
  cross-correlation function between dark  matter haloes and  the mass
  distribution  within   their  domains.  We   use N-body cosmological
  simulations  to   show that  the  density  profiles  required in our
  reconstruction scheme   can  be  determined  reliably   from   large
  cosmological simulations,  and that our  method  can reconstruct the
  density field accurately  using  haloes  with masses down   to $\sim
  10^{12}\msun$  (above which  samples    of  galaxy groups  can    be
  constructed   from current   large  redshift surveys  of  galaxies).
  Working in   redshift  space, we   demonstrate that    the redshift
  distortions  due   to  the  peculiar  velocities  of haloes  can  be
  corrected in an  iterative way.  We  also describe some applications
  of our method.
\end{abstract}


\section{Introduction}

There is now  much evidence that we  live in a flat universe dominated
by cold dark matter (CDM, $\sim 30\%$) and  dark energy ($\sim 70\%$). In
addition, perturbations in the density  field are characterized by  an
initial  power spectrum   with a  spectral   index $n \sim  1$,  and a
normalization $\sigma_8 \sim 0.8$. Here $\sigma_8$ is the {\it RMS} of
the linear perturbation field at   the present, in spheres of  8$\mpc$
radius.  This `standard'    $\Lambda\rm{CDM}$  model   has  been  very
successful   in explaining   a variety   of   observations,  such   as
temperature fluctuations in    the cosmic microwave   background,  the
clustering  of  galaxies on large   scales, and the clustering  of the
Lyman-$\alpha$ forest at  high redshift (e.g.  Spergel et al, 2007 and
references  therein). In  the  CDM  cosmogony, a   key  concept in the
build-up  of structure is the  formation of dark matter haloes, formed
through non-linear gravitational  collapse. In a hierarchical scenario
like CDM,  most of the  mass  at any  given time  is bound within dark
haloes of various  masses;  galaxies and  other luminous   objects are
assumed to  form by cooling  and condensation  of baryonic gas  within
these  haloes (e.g.   White  \&   Rees  1978).   With current   N-body
simulations  and  analytic  models, the  properties  of   the CDM halo
population, such  as the  mass function,  the spatial  clustering, the
formation history  and  the  internal structure  are well  understood.
However, the details of how galaxies form in  the cosmic density field
are still poorly understood.

A key step  in understanding  galaxy  formation in the cosmic  density
field is to understand the relationships between galaxies, dark matter
haloes and  the large scale  structure.  The study of  the large-scale
structure in  the universe typically relies  on large redshift surveys
of galaxies,  such as the Sloan  Digital  Sky Survey  (hereafter SDSS,
e.g.    York et al.   2000)  and  the  2-degree Galaxy Redshift Survey
(2dFGRS, e.g. Colless et al. 2001).   However galaxies are known to be
biased tracers of  the large scale mass  distribution.
Unfortunately,  the  exact   form of  this  bias   is
complicated, as it depends on various properties of the galaxies, such
as luminosity and color.  Gravitational lensing and X-ray observations
may be used to  probe  the  mass  distribution in  individual  massive
systems (see  the review   of Bartelmann   \& Schneider  2001;  Evrard,
Metzler, Navarro 1996; Gastaldello et al. 2007), such as clusters, but
the  majority of  the mass,  which  is contained  in  systems of lower
masses,  cannot be  probed  in this   way.  On super-cluster   scales,
gravitational weak lensing can be used to study  the properties of the
mass  distribution in a statistical way,  but it is still difficult to
use  this  method to directly identify    the large-scale structure in
detail.

One  important development in recent  years is that tremendous
amounts of  effort have been put   into the establishment  of the
relationship between galaxies and dark matter  haloes, using
numerical  simulations (e.g., Katz, Weinberg \& Hernquist  1996;
Pearce et al. 2000; Springel 2005; Springel et al.  2005) or
semi-analytical models (e.g.  White \& Frenk 1991; Kauffmann  et
al. 1993,  2004; Somerville \& Primack 1999; Cole et al. 2000; van
den Bosch 2002; Kang  et al. 2005; Croton et al. 2006). Both of
these techniques  try to model   the process of  galaxy formation
ab initio.  However, since our  understanding of the various
physical  processes involved is  still  relatively poor, the
relations between the properties  of   galaxies and  their dark
matter  haloes predicted  by these simulations and semi-analytical
models still need to be tested against observations. More
recently, the halo occupation model has opened another avenue  to
probe the galaxy-dark matter halo connection (e.g.  Jing, Mo \&
B\"orner 1998;  Peacock \& Smith 2000; Berlind \& Weinberg  2002;
Cooray \& Sheth  2002; Scranton 2003; Yang, Mo \&  van den  Bosch
2003;  van  den Bosch,   Yang \& Mo   2003; Yan, Madgwick \& White
2003; Tinker et al.  2005; Zheng et al. 2005; Cooray 2006;  Vale
\&  Ostriker  2006;  van den Bosch et  al.   2007).  This
technique uses the  observed galaxy luminosity function and
two-point correlation functions to constrain the average  number
of galaxies of given   properties that occupy  a  dark  matter
halo   of  given mass. Although  this method  has the  advantage
that it  can yield much better  fits to the  data than   the
semi-analytical models   or numerical simulations, one typically
needs to assume a somewhat ad-hoc functional form  to describe the
halo occupation model. A more direct way of studying the
galaxy-halo connection is  by using galaxy groups, provided that
these are defined as sets of galaxies that reside in the same dark
matter  halo. Recently,  Yang    et al. (2005; 2007)  have
developed  a halo-based group  finder  that is  optimized for
grouping galaxies that reside in the same dark matter halo.  Using
mock galaxy redshift surveys  constructed from the conditional
luminosity function model (e.g. Yang,   Mo \& van den Bosch 2003)
and a  semi-analytical model (Kang et al, 2005), it is found that
this group finder is very successful in associating galaxies  with
their common   dark  matter haloes (see Yang et al. 2007). The
group finder also performs reliably for  poor systems, including
isolated  galaxies in small mass haloes, making  it ideally suited
for  the study  of  the relation between galaxies and dark matter
haloes over a wide range of halo masses. Thus far, this halo-based
group finder has  been applied to 2dFGRS (Yang et al.  2005), SDSS
DR2 (Weinmann et al. 2006) and  SDSS DR4 (Yang et al. 2007).

In the current CDM model, the relationship between dark haloes and the
mass density field can be understood using $N$-body simulations and/or
analytical models. This relationship,  together with the galaxy groups
that represent dark haloes,  offers the possibility to reconstruct the
underlying  cosmic density field  with   the use  of the  dark  haloes
represented by galaxy systems. If this approach proves successful, its
applications to real  observations will enable us  to map the  density
field in  the local universe,  allowing   us to  study  in detail  the
relationships between galaxies, dark haloes and large-scale structure.
In this paper  we develop a method to  reconstruct  the cosmic density
field from the distribution of dark matter haloes. Our method uses the
fact that the statistical properties of  the large-scale structure are
well represented  by the current   $\Lambda$CDM  model and that   dark
haloes can   be  selected  reliably from   large  redshift surveys  of
galaxies, as mentioned above.

The  reconstruction    of  the  cosmic  density    field   from
galaxy distribution  has been carried out   earlier based on
various redshift surveys (e.g. Fisher et al. 1995; Zaroubi et al.
1995; Schmoldt et al. 1999;  Mathis et  al.  2002; Erdo{\u  g}du
et  al.  2004).  In  these investigations, the   distribution of
galaxies  is  usually  smoothed heavily and normalized to
represent the  cosmic density field on large scales.  In the
Wiener  reconstruction method  adopted  in many of the earlier
investigations,  the mass density at a  given point is assumed to
be  a  linear combination of the observed galaxy density
field values at different points so that
the reconstructed field has the   minimum mean square error. Our
method is different from these methods in that it is based on dark
matter haloes  so that  the reconstruction  is more  accurate on
small scales.  Furthermore, since our method  is based on dark
matter haloes represented   by galaxy systems,   the bias  of the
distributions of different  galaxies relative   to the  underlying
density field    is automatically  taken into account by  their
connections to dark matter haloes.

This paper is arranged as follows. We describe briefly the simulations
to be  used    and  how  dark   haloes  are   identified in    Section
\ref{sec_sim}. In  Section \ref{sec_ccf} we present our reconstruction
method and calculate the density  profiles within and around haloes of
different masses.  In  Section \ref{sec_rec}  we  compare  our density
field   reconstructed from   the   halo  catalogue  selected from  the
simulations with the original  density field. In Section \ref{sec_red}
we examine how our reconstruction  scheme works in redshift space.  In
Section \ref{sec_app}  we outline some  potential  applications of our
method, and we summarize our results in Section \ref{sec_sum}.

\section{Simulations}\label{sec_sim}

In this paper, we use two sets of $N$-body simulations and dark
matter haloes  selected  from them to   test  the  reconstruction
method  we propose.   Here we give a  brief description of these
two simulations. These simulations  are obtained using  the ${\rm
P^3M}$ code described in Jing \& Suto (2002). The main simulation,
which will be referred to as L300, assumes  a  spatially-flat
$\Lambda$CDM model,   with density parameters $\Omega_{\rm m}=0.3$
and $\Omega_\Lambda=0.7$, and with  the CDM power spectrum given
by Bardeen et al (1986),  with a shape parameter
$\Gamma=\Omega_{\rm m}h=0.2$ and an  amplitude specified  by
$\sigma_8=0.9$. The CDM density field was traced with $512^3$
particles, each having a mass  of  $M_p\sim1.68 \times
10^{10}$$\msun$,  in a  cubic box of 300 $\mpc$. The softening
length is  $\sim 30 \kpc$. The other simulation, referred to as
L100 in the  following, assumes the  same cosmological model as
L300,  and uses the   same number   of  particles,  but the
simulation  box is   smaller,  100$\mpc$, and the   mass
resolution is higher, $M_p\sim6.2\times10^{8}\msun$.

Dark matter haloes were identified with a friends-of-friends algorithm
with a link  length that is 0.2  times the  mean particles separation.
The mass of a halo, $M_h$, is the sum of the mass of all the particles
in the halo. The virial radius $R_h$ of a halo is defined as:
\begin{equation}
\label{eq_rvir} R_h=\left({3 M_h\over 4\pi \Delta_h{\rho_{\rm
m}}}\right)^{1/3}\,,
\end{equation}
where   $\rho_{\rm m}$  is  the  mean mass   density  of  the
universe, and $\Delta_h$ is the  mean  density contrast  of  a
virialized  halo.  We choose  $\Delta_h =   200$, but the   exact
choice does   not  have a significant impact on our results.

\begin{figure}
\epsscale{1}\plotone{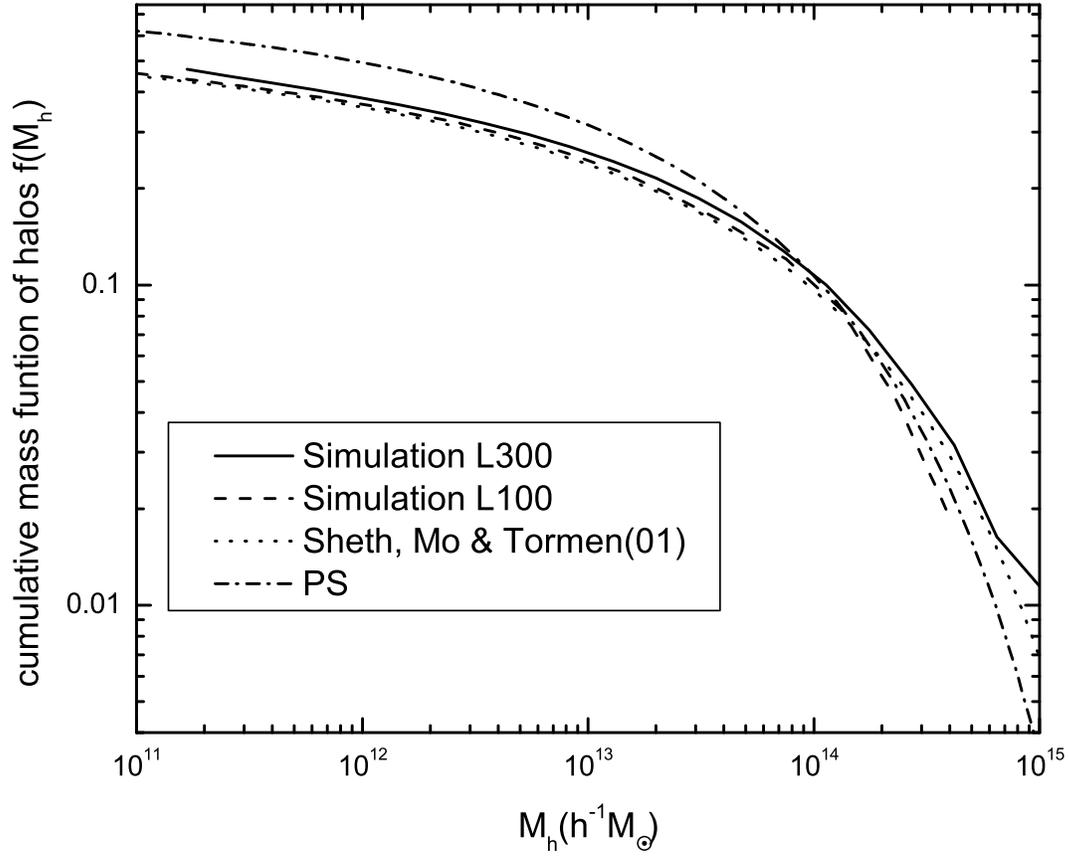} \caption{The fraction of cosmic mass
contained in haloes with mass above $M_h$ as a function of
$M_h$.}\label{fig_mhf}
\end{figure}

In Fig.  \ref{fig_mhf}, we show the  fraction of cosmic mass contained
in haloes more  massive than $M_h$ as  a function of  $M_h$. Note that
the mass functions given by the two simulations  are similar except at
the  massive end where the  small-box simulation gives a significantly
lower   fraction. This is largely  due   to the  box-size effect.  For
comparison, we also plot  the predictions based on the Press-Schechter
mass function  (Press  \& Schechter  1974)  and the mass   function of
Sheth, Mo \& Tormen (2001).  As one can  see, about 50\% of the cosmic
mass is contained in   haloes with masses larger than  $10^{11}\msun$,
and about 40\%  is in haloes more  massive than $10^{12}$  $\msun$, in
the CDM model considered here. If the normalization, $\sigma_8$, has a
lower value, as may  be the case according to   the recent WMAP3  data
(e.g.  Spergel et al. 2007), these fractions are even lower.  As shown
in Yang et al.  (2007), current large  redshift  surveys, such  as the
SDSS, can  be   used  to select   galaxy  groups with     masses
down to
$10^{12}\msun$. Thus, more than $\sim 60$\%  of the cosmic mass is not
directly associated  with the viralized   haloes of the  galaxy groups
that can be reliably identified from current  redshift surveys.  It is
therefore very  important  to investigate whether the   mass component
that is {\it not} directly   observable can be reconstructed based  on
what we can see.

\section{Quantifying the mass distribution in and around dark matter haloes}
\label{sec_ccf}

\subsection{The partitioning of the mass distribution}\label{sec_domain}

The  goal of  this paper is  to use   the distribution  of dark matter
haloes to reconstruct the  cosmic density field. Since observationally
we can only identify haloes  above some mass  threshold, the method is
useful only  if such reconstruction is  based on haloes massive enough
to    be observationally identifiable.    For  a  given mass threshold
$M_{\rm th}$,  we refer all haloes with   masses above it as  the halo
population.  The cosmic mass  (dark matter particles) in the  universe
can then be divided  into  two parts:  (i)  the halo component,  which
contains all the particles that  are assigned to  the proper of haloes
above  the mass  threshold; (ii)  the  complementary  component, which
contains  particles  that are not    assigned to the   proper of these
haloes. Note that some of the particles in the complementary component
are in haloes of lower masses, while some may be in a diffuse form. As
an   illustration,  we  show  in   Fig.~\ref{fig_md}   the  total mass
distribution  in  a  slice  of the  L300   simulation, along with  the
distributions  of  the  halo   and  complementary components, assuming
$M_{\rm th}=10^{12}\msun$.  Clearly, the large-scale structure is well
traced by  the halo   component,  and the complementary  component  is
tightly correlated with the halo component.   Therefore, it is hopeful
to use the halo component to reconstruct the full density field.

For a  given halo population,  we  can partition  space into a  set of
domains, each  of which contains  one halo. The domain  of any halo is
defined in such a way that  each point in the  domain is closer (based
on  a distance measure  to be defined below) to   the halo than to any
other  haloes in  the halo population.    With a proper  definition of
distance measure, each volume element (and  each mass particle) can be
assigned uniquely to a domain.  The particles in the  domain of a halo
will be referred to as the domain particles of the halo.

We use the following quantity to describe  the proximity of a point to
a halo of virial radius $R_h$:
\begin{equation}
r_{d}=\frac{r_h}{R_h}\,,\label{eq_rd}
\end{equation}
where $r_h$ is the  physical distance between  the halo center and
the point. As an illustration, Fig.~\ref{fig_domain} shows the
domains of 10 randomly distributed  haloes. Note that some small
haloes and their  domains are embedded in the domains of larger
haloes.  In this way, each particle in  the complementary
component is assigned to a unique domain. In the next section, we
will discuss why we choose to use $r_d=r/R_h$ to define domains.

\subsection {Mass distribution in and around haloes}

In order to  reconstruct the cosmic  density field accurately, we need
to model the mass distribution in  and around dark matter haloes. Here
we use the cross-correlation function between  the haloes and the mass
in  their domains to characterize the  relation between the haloes and
the surrounding mass   distribution.   This  is different  from    the
conventional cross-correlation function, which  is defined by the mass
density within spherical shells of a given radius  centered on a halo,
regardless whether or not the particles are in the domain of the halo.
The  advantage of our definition is  that each particle (mass element)
contributes  only to the cross-correlation with  one dark matter halo,
thus avoiding  the  problem  of  double-counting  when  we  use  these
profiles to  reconstruct the density  field.  This turns  out to be an
important advantage.  For instance, in the conventional definition, if
we  use the cross-correlation function  as the average density profile
around a halo to reconstruct the density field, then the reconstructed
field will contain more mass than the original  field, which has to be
corrected  for.  However, it  is  not straightforward  to rescale  the
density field   while  preserving the density    fluctuations on small
scales.   With our  domain-based definition  of  the cross correlation
function, however, mass conservation is guaranteed by construction.

Since we want to  obtain the correlation between  the haloes and their
corresponding  domain particles,  we modify   the  definition of   the
cross-correlation function as follows.  First,  for a given halo and a
given spherical    shell  (with  radius between   $r-\delta  r/2$  and
$r+\delta r/2$ in unit of virial radius)  centered on it, we calculate
the volume of  the intersection between  its domain  and the spherical
shell [which we denote by $V_h(r)$] as
\begin{equation}\label{eq_vhr}
V_h(r)=\sum_{p=1}^{N_h^p(r)}V_p\,,
\end{equation}
where  $N_h^p(r)$ is the number of  domain  particles within the shell
and  $V_p$  is  the effective  volume  of  the Delaunay  tessellations
associated  with a  domain particle  $p$.  Direct  calculation of  the
intersection volume is very time consuming. This  is the reason why we
use  instead the estimate    given  by equation  (\ref{eq_vhr})   (see
Appendix A for details).  Our test based on a  subset of haloes showed
that this estimate is sufficiently accurate for the calculation of the
cross correlation function. Thus, for a set of $N_h$ haloes, the total
volume  of such intersections can  be obtained by  summing up $V_h(r)$
over all the haloes. We denote this volume by $V_{s}(r)$. Finally, the
average density profile can be estimated from
\begin{equation}\label{eq_srho}
\bar{\rho}(r)= \frac{\sum_{h=1}^{N_h}\sum_{p=1}^{N_h^p(r)}
M_p}{V_s(r)} =\frac{\sum_{h=1}^{N_h}N_h^p(r) M_p}{\sum_{h=1}^{N_h}
\sum_{p=1}^{N_h^p(r)}V_p}\,,
\end{equation}
where $M_p$ is  the mass of a particle.  As mentioned  above, this
average density  profile  is   different   from  that  based on
the conventional  cross correlation   function  between haloes and
mass, because  here the  number of  halo-particle pairs  is
estimated  only between a halo and its domain particles. In Fig.
\ref{fig_s11} we show the averaged density profiles  for haloes in
6 mass bins, using  $M_{\rm th}=1.68\times10^{11}\msun$.
Note that the mass resolution of the L300 simulation is
insufficient to resolve small halos, and it is used only to
derive the density profiles in the domain associated with halos
above $10^{13}\msun$, which contain at least 590 particles.
For smaller halos, the profiles are obtained from L100.
The boxsize of L100 may be too small to represent a real survey,
but should be reliable for estimating the profiles around halos.
Note also that, since the purpose of the paper is to test the
reconstruction method, rather than probing the model prediction,
the inaccuracy of the simulation should not have significant
impact on our reconstruction which involves other more serious
uncertainties (see below).

For comparison,    the   corresponding  results  using $M_{\rm
  th}=10^{12}\msun$ and $M_{\rm th}=10^{12.5}\msun$ are plotted in
Figs.  \ref{fig_s12}  and \ref{fig_s12.5},  respectively. As shown
in Yang et al.  (2005;  2007), the  values $M_{\rm
th}=10^{12}\msun$ ($3.16\times10^{12}\msun$) represent the
lower-mass  limit to  which a complete group sample can be
selected from the SDSS galaxy  catalogue to  a redshift of  $z\sim
0.1$ (0.15). The density profiles within the  virial radius are
well fitted  by NFW profiles (Navarro et al. 1997), and the
resulting concentration-mass relation  is in a good agreement with
that obtained by, e.g.  Bullock et al. (2001). Here we do not take
into account the scatter in the concentration-mass relation,
because it is difficult to obtain the concentrations of individual
systems from observation and it is important to examine how the
lack of such information affects the reconstruction results. At
radii larger than $R_h$, the profiles we obtain are comparable to
what Prada et al. (2006) obtained for  isolated  haloes.  As one
can see, the density profile is measured over the range from 0.05
$R_h$ to about 30 $R_h$, or in terms of density, from $\sim
4\times 10^4\rho_{\rm m}$ to $\sim 0.1\rho_{\rm m}$.  These
demonstrate that, for   a given cosmology, current computer
simulations can be used   to determine   reliably   the
cross-correlation functions to be  used in our  reconstruction
model. The average density profile around $(2\to 3) R_h$ decreases
slightly with increasing halo mass, mainly because the effect of
infall increases with halo mass (e.g. Prada et al. 2006). The
density profile at large scales is  higher for  a  higher mass
threshold, because a large fraction of the cosmic mass is assigned
to the complementary component.

In what follows, we will use these density profiles and
halo/domain information to make the reconstruction.
The choice of the definition of $r_d$ is only useful
if the derived profiles as function of $r_d$ depend
only weakly on halo mass so that the reconstructed density
has no jump at the boundary of the domains. The cross
correlation within the virial radius has the same form
as the NFW profile, which depends only weakly on halo
mass when the radius is scaled with the halo virial radius.
As shown in Figs.\ref{fig_s11} -- \ref{fig_s12.5},
the density profiles outside the virial radius also
depend weakly on halo mass, indicating the usefulness of
using $r_d$ to represent radius.

\section{Reconstructing the cosmic density field}
\label{sec_rec}

As we have mentioned in the introduction, our reconstruction method is
based  on the assumption  that  the density  field is  similar to that
predicted by the current $\Lambda$CDM model, and we  have shown in the
previous section that  the density profiles in  and around dark matter
haloes, which are required in our reconstruction model to be described
below,  can be estimated reliably  from current numerical simulations.
In this  section we will  describe how to  reconstruct the dark matter
distribution based  on the distribution of dark  haloes. We will first
introduce  our  method and do  the   reconstruction based on  the halo
population in simulation L300,  although the density profiles are from
both L300   and L100. Then  we compare  our results with  the original
simulation, i.e. L300, by using several methods.

It should be pointed out that, although our reconstruction is based on
the two-point correlation function between haloes and the mass density
field, it contains high-order information through the inclusion of the
halo-halo correlation, and of  some environmental information, such as
that represented by the domains.

\subsection{The Reconstruction Method}

Our reconstruction scheme consists of the following steps.

\begin{itemize}
\item We start with   a sample of  dark  matter  haloes above a   mass
  threshold, $M_{\rm th}$, each of which has a mass and a position
  in space.
\item For a  halo $h$  we pick  the average  density profile for   all
  haloes  in a given mass  bin that includes the mass  of  the halo in
  question.
\item We use a Monte-Carlo  method to put particles around the
  halo $h$ up to  $\sim 30$  times virial radius
  (sufficient to cover the domain of the halo) regardless of the
  domain, using the density  profile picked above.  The mass of
  each of the sampling particles is exactly the same as
  the particle mass in the simulation  L300.
  For convenience, we use $S_h$ to denote this set of sampling
  particles.
\item  For each particle in $S_h$, we check whether the distance
  $r_d$ [defined in equation (\ref{eq_rd})] between the particle
  and the center of halo $h$ is smaller than  that  to any  other
  haloes in the halo population.  If yes, the
  particle is retained; otherwise it is removed.
  Thus, only particles in the domain  of halo $h$ are retained.
\item Repeating the Steps 2, 3, and 4 for  all haloes in the halo
  population, we obtain a reconstructed density field sampled by
  particles. Because of the way our cross-correlation function
  between halos and domain particles are defined, the reconstructed
  density field has a total mass that is very similar to the
  original field.
\end{itemize}

On large scales, our  reconstruction  method ensures that the  density
field is   the same as  that  traced by  the distribution of  the halo
population.    On   small     scales,  our    method reproduces    the
cross-correlation between haloes and dark matter.  The idea behind our
method is very  similar to that in the  current halo model (e.g.  Jing
et  al.  1998; Cooray  \&  Sheth 2002  and references  therein), which
models the mass distribution   by convolving the distribution of  dark
matter   haloes with their density   profiles. The difference is that,
while the halo model is based on all haloes and their density profiles
within the  virial radii, our model is  based on  relatively high mass
haloes and the distribution of the complementary mass component around
these haloes. Our approach is more useful in reconstructing the cosmic
mass  density field from  observations,  because redshift surveys  can
only  be  used to  identify  relatively  massive haloes,  as described
above.

\subsection{The reconstructed versus the original density
fields}\label{sec_comd}

To compare  our reconstructed density field  with the original
density field, we use the  SPH method (Monaghan 1992) to  smooth
the reconstructed and simulated particle  distributions on a
Cartesian grid with grid-sizes of $l_{\rm  b}  = 250  \kpc$ (see
Appendix B for the detail). In what follows, we refer to this
grid as our `base-grid'. In order to investigate the accuracy of
our reconstruction method, we compare the  reconstructed density
field  $\rho_{\rm rec}$   and  the original density field
$\rho_{\rm   sim}$  by  applying  different smoothing  methods and
smoothing  scales on the density field on the base-grid.

The top-left panels of Fig. \ref{fig_s11_dc}, \ref{fig_s12_dc} and
\ref{fig_s125_dc} show    the comparison  in which  the mass
distributions   on   our base-grid are box-car smoothed on a scale
of  $l_{\rm sm}=1\mpc$ for three different reconstructions  with
halo  mass  thresholds  of $1.68\times10^{11}$, $10^{12}$ and
$10^{12.5}\msun$, respectively. Note that the smoothed density is
simply  computed by averaging over the enclosed  base-grid cells,
which  in the  case of $l_{\rm sm} =   1 \mpc$  corresponds to
$28^3$ grid-cells.  The solid line represents the mean relation,
while the error bars indicate the standard deviation in $\rho_{\rm
sim}$ for given $\rho_{\rm rec}$.  As one can see, the bias in the
mean relation is very small. However, in the low-density regime
the scatter is quite large, especially for reconstructions  with
large $M_{\rm th}$.   This is expected, because  the density field
in the low-density regions is not sampled in  detail by the
population of massive haloes.  In order to suppress these
fluctuations, one may smooth the density field  on even  larger
scales.  In the middle-left   and bottom-left panels, we show
results using $l_{\rm sm}=2\mpc$  and $4\mpc$, respectively.   As
expected, the scatter in the low density bins is now reduced.

Unfortunately,  the  use  of  a large   smoothing  length  dilutes the
high-density regions, thus reducing  the dynamical range probed.  As a
compromise,   we therefore introduce   an  adaptive  smoothing  length
$l_{\rm ad}(M_{\rm s})=n l_{\rm b}$ to smooth the density field, where
$M_{\rm s}$ is a chosen smoothing mass scale (hereafter referred to as
SMS), and $n$  is an adjustable,  even  integer.  The value of  $n$ is
tuned  so that the  mass    contained in  the super-grid of    $l_{\rm
  ad}(M_{\rm s})$ first reaches $M_{\rm  s}$.  In the right panels of
Fig.   \ref{fig_s11_dc},  \ref{fig_s12_dc} and \ref{fig_s125_dc},
we show the comparison between the reconstruction and simulation
using $M_{\rm s}=M_{\rm  th}/2$, $M_{\rm th}$  $2M_{\rm th}$ from
top  to bottom.  Note that it is possible that at  a given
location the value of  $l_{\rm ad}(M_{\rm s})$   obtained from the
reconstructed  density field is different from that obtained from
the original density field in the simulation.  The adaptive
smoothing length used above is based on the  reconstructed field.
By using $M_{\rm s}=1\sim2M_{\rm th}$, our method reliably
reconstructs the density field over a very large density range,
because the small structures represented by halos with masses
below $M_{\rm th}$, which cannot be recovered by our method, is
effectively smoothed. The large scatter in the intermediate
density bins, $\sim 20\rho_m$, in the top-right panel of Fig.
\ref{fig_s11_dc} ($M_{\rm th}=1.68\times10^{11}\msun, M_{\rm s}
=M_{\rm th}/2$) is probably due to the neglect of substructures,
especially large subhalos, in dark matter halos.
Another reason is the definition of FOF halos, which
suffer from the halo bridging problem (Tinker et al. 2008,
Lukic et al. 2008). However, as discussed in Section 4.4,
the use of elliptical model, which takes care of the `bridging'
problem to some degree, does not make a significant improvement.
Our test using spherical overdensity halos did not make
significant difference.

As  an   illustration    of   the quality of      our
reconstruction, Fig.~\ref{fig_lsscon} shows the   contours of  the
projected  density distribution in  a  slice $120 \times   120
\times 10  (\mpc)$  of the reconstructed field (middle  panels) in
comparison with those obtained from the original density field
(left panels). The different rows show contours at different
levels. The  mass threshold adopted here is $M_{\rm
th}=10^{12}\msun$, and the   density fields are smoothed  on  a
scale of   $l_{\rm  sm}=4\mpc$. As one can see, the
reconstructed density field shows filamentary structure
connecting high density nodes, similar to that in the original
field. This is similar to the quantitative comparisons
shown presented above.

We  have   also computed  the   power spectrum  of  the  original
and reconstructed density fields,  using  a Fast Fourier Transform
of the overdensity on a  $1024^3$ grid   (corresponding   to an
effective smoothing scale  of $l_{\rm sm}  \sim 300 \kpc$). The
overdensity is obtained using  the Cloud-In-Cell (CIC) weighting
scheme (Hockney \& Eastwood   1981).  The resulting power  spectra
are shown  in the top panel of Fig \ref{fig_pk} in comparison with
that obtained directly from the simulation data (solid line). The
bottom panel shows the difference between the reconstructed and
original power spectra  as a function of wave number $k$. The
reconstruction results match the simulation very well on large
scale. The fractional difference at $k<3\mpci$ is less than 14\%
for all the reconstructions.  The discrepancy is believed
to be generated by the inaccuracy of our simple model for
the mass distribution. For example, the underestimation on small
scales ($k>3\mpci$) is almost certainly due to the neglect of the
structure in the complementary component produced by  low mass haloes,
the neglect of the substructure in massive haloes, and
the bridging effect in FOF halos.

\subsection{The predicted versus the original velocity fields}
\label{sec_vel}

Since our final goal  is to apply our method  to observational data in
redshift space, we have to deal with redshift space distortions due to
the peculiar velocities of  dark matter haloes.  It  is known that the
large-scale  peculiar velocity  field is strongly  correlated with the
gravitational acceleration field (e.g.   Colombi et al.  2007), and so
it may  be possible to predict   the current velocity field  using the
current density field.  If this is the case,  we can then start with a
density field in the redshift space, make  corrections to the peculiar
velocities, and iterate  to  get the real-space density  distribution.
In this subsection,  we study how the  reconstructed density field can
be used to make predictions for the peculiar  velocities of haloes. In
the next section, we will show how one can use this information in our
reconstruction in order to account for redshift space distortions.

In the linear regime, the peculiar velocity is  related to the density
perturbation as
\begin{equation}\label{eq_hv}
\textbf{v}(\textbf{k})=Hf(\Omega)
\frac{i\textbf{k}}{k^2}\delta_{\textbf{k}}\,,\label{eq_vk}
\end{equation}
where  $\textbf{v}(\textbf{k})$    and  $\delta_{\textbf{k}}$ are
the Fourier  transforms of the velocity  field  and mass density
contrast, respectively,       $H$    is       the      Hubble
constant   and $f(\Omega)=\Omega_{\rm
m}^{0.6}+\Omega_{\Lambda}/70(1+\Omega_{\rm m}/2)$ (e.g. Lahav  et
al. 1991).  Note    that we are  estimating the peculiar
velocities of virialized   haloes,  for  which  the linear   model
is expected to yield a reasonable approximation. Since our
reconstruction method is halo-based,  the (strongly non-linear)
virial motions within dark matter haloes are irrelevant.

We use the CIC  scheme to construct the  density field, on  a
$1024^3$ grid, from the particles representing the reconstructed
density field. Because the   sizes of different  haloes are
different, we divide the halo population into  6 subsamples
according to   their masses. For  a given subsample of haloes we
use  top-hat windows with radius equal to the average Lagrangian
radius of the haloes  in the sample to  smooth the   density
field.    We  then  use    Fourier transform  to  obtain
$\delta_{\textbf{k}}$,  and use  equation (\ref{eq_hv})  to obtain
the linear velocity  field.  We use the linear   velocity
predicted at the position  of a halo   to  represent  the
predicted velocity of   the corresponding  halo.  In Fig.
\ref{fig_vel}    we show the $x$-component of the predicted velocity
versus that of the true velocity,  using $M_{\rm
th}=10^{12}\msun$. Overall, the predicted velocity is tightly
correlated with  the true velocity. For a very small fraction of
haloes, the predicted velocity is significantly larger than  the
true one.   These haloes are usually close   to  massive  haloes,
so that non-linear effects cannot be neglected, rendering Equation
(\ref{eq_vk}) inaccurate.

The   dashed line   in   Fig.~\ref{fig_vd}    shows  the   probability
distribution of the difference between the predicted velocity based on
the  reconstruction and the  true velocity.  For comparison, the solid
lines  shows the  same distribution  but  obtained using the simulated
mass distribution.  The two distributions are very similar, indicating
that most of the error in the peculiar velocities is not due to errors
in the    reconstructed  density  field,  but  due   to   the  limited
applicability   of linear theory.  Our result    is similar to that of
Lavaux (2008), who  used the  Monge-Ampere-Kantorovitch method
to infer the peculiar velocities of haloes and galaxies.

\subsection{Spherical versus Elliptical Model}

\begin{figure}
\epsscale{0.5}\plotone{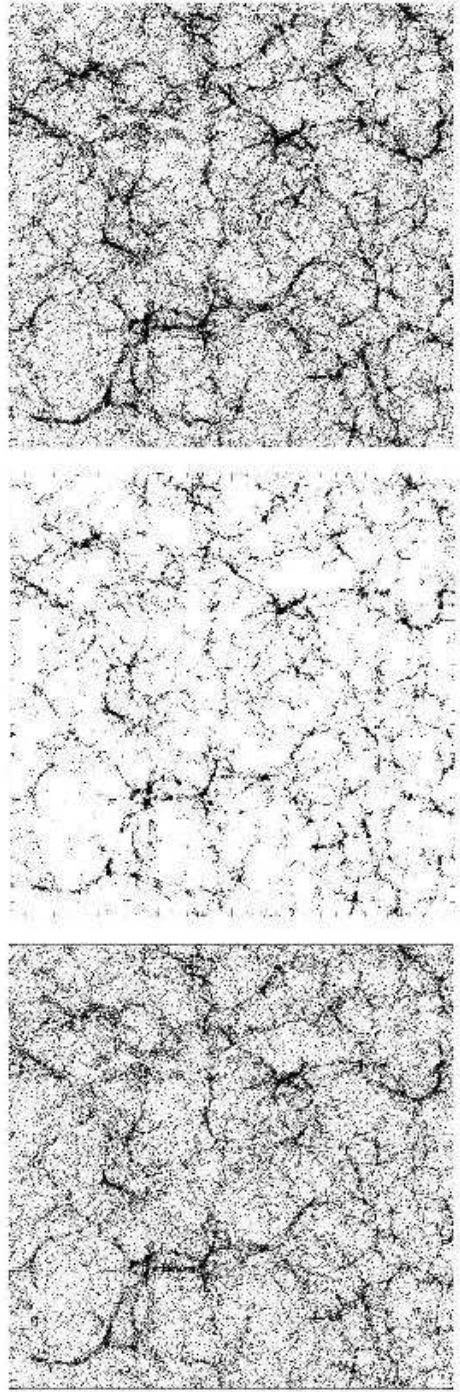} \caption{The distribution of all
mass (upper panel), the mass in the halo population (middle panel)
and the mass not in the halo population (lower panel) in a
simulation slice of $300\times 300 \times 10 (\mpc)^3$. The
haloes, shown here, have masses larger than
$10^{12}\msun$.}\label{fig_md}
\end{figure}

\begin{figure}
\epsscale{1}\plotone{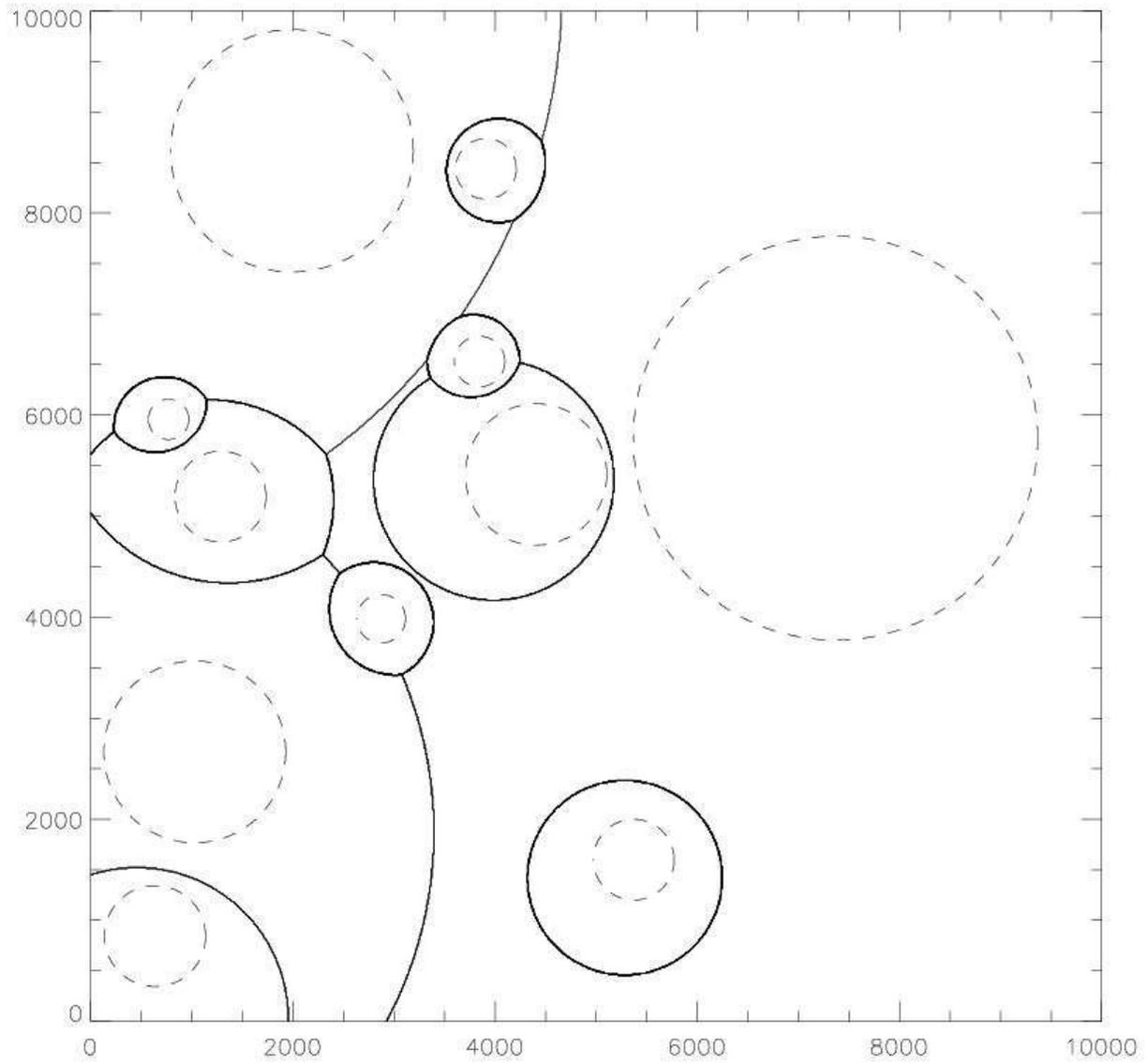} \caption{The domain of randomly
distributed haloes. The dashed lines represent the haloes, scaled
by the virial radius of each halo. The solid lines are the
boundaries of the domains (see text for
details).}\label{fig_domain}
\end{figure}

\begin{figure}
\epsscale{1}\plotone{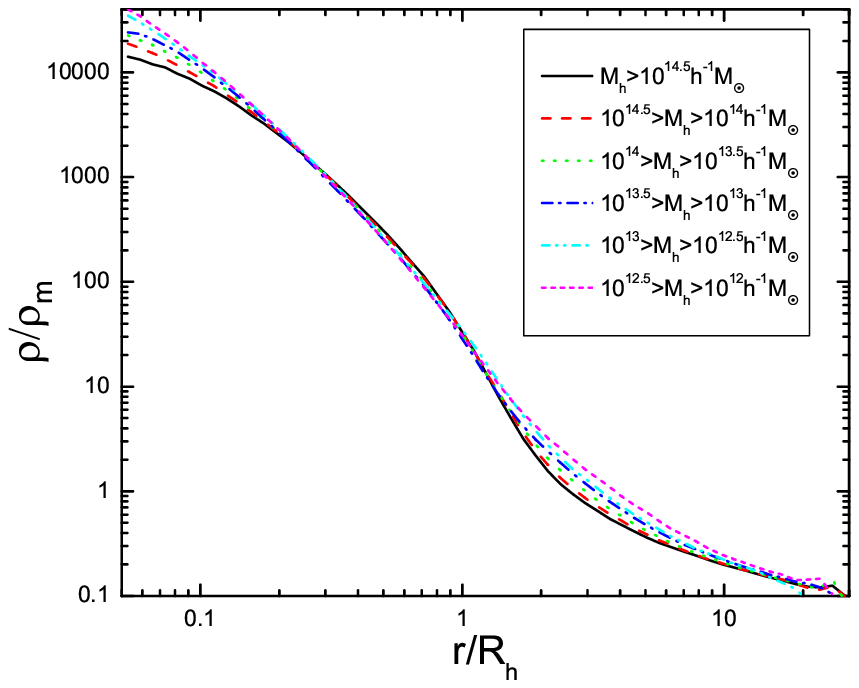} \caption{The density profiles of mass
in and around the haloes in various mass bins. Here the mass
threshold for the halo population is $M_{\rm
th}=1.68\times10^{11}\msun$. The radius $r$ is scaled by halo
virial radius $R_h$, and the density is scaled with $\rho_m$, the
mean density of the universe.}\label{fig_s11}
\end{figure}

\begin{figure}
\epsscale{1}\plotone{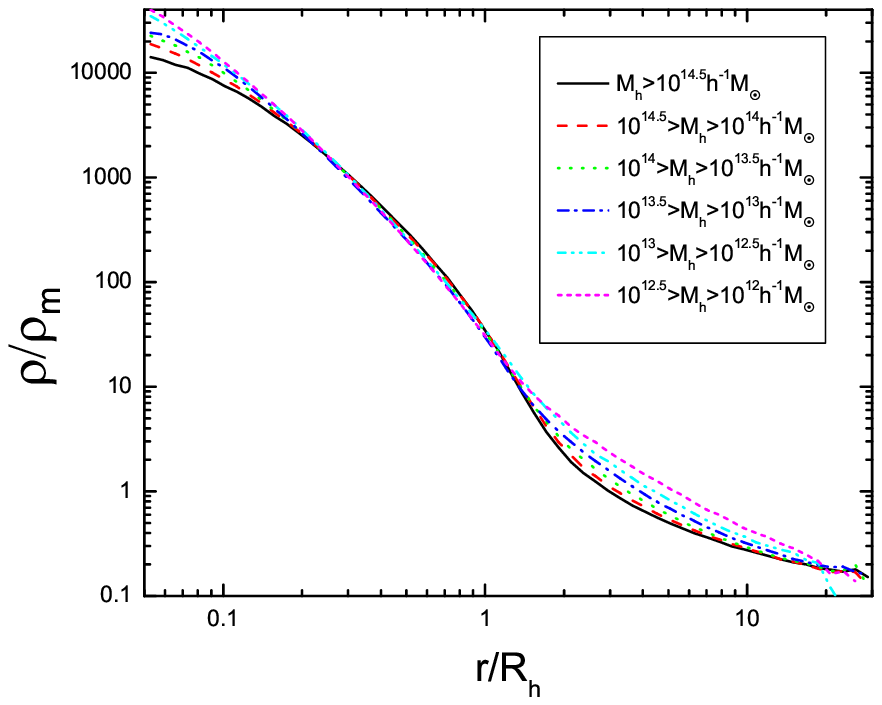} \caption{The density profiles of mass
in and around the haloes in various mass bins. Here the mass
threshold for the halo population is $M_{\rm
th}=1.0\times10^{12}\msun$. The radius $r$ is scaled by halo
virial radius $R_h$, and the density is scaled with $\rho_m$, the
mean density of the universe.}\label{fig_s12}
\end{figure}

\begin{figure}
\epsscale{1}\plotone{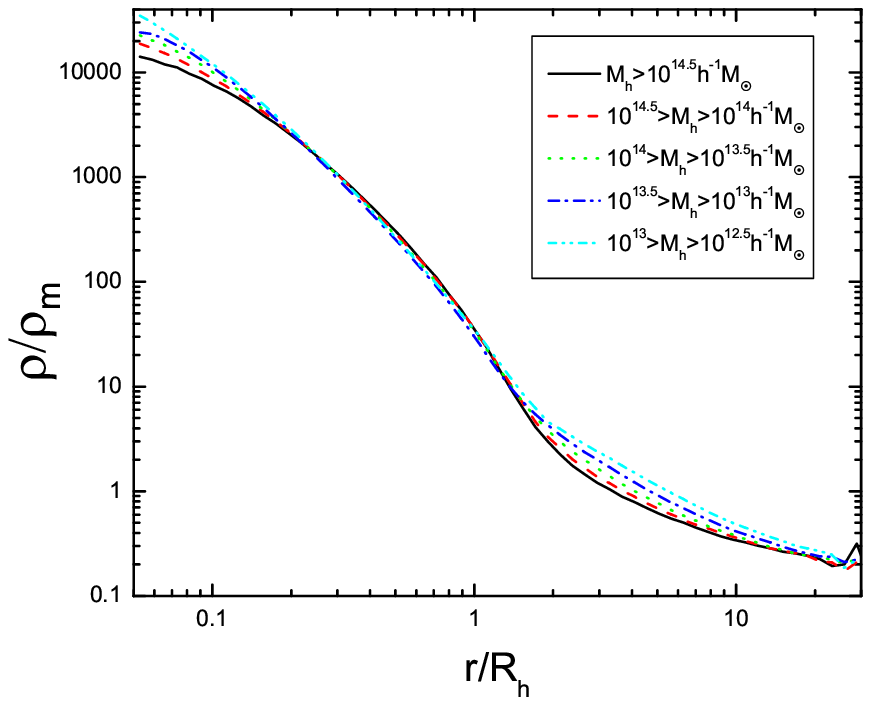} \caption{The density profiles of mass
in and around the haloes in various mass bins. Here the mass
threshold for the halo population is $M_{\rm
th}=1.0\times10^{12.5}\msun$. The radius $r$ is scaled by halo
virial radius $R_h$, and the density is scaled with $\rho_m$, the
mean density of the universe.}\label{fig_s12.5}
\end{figure}

\begin{figure}
\epsscale{1}\plotone{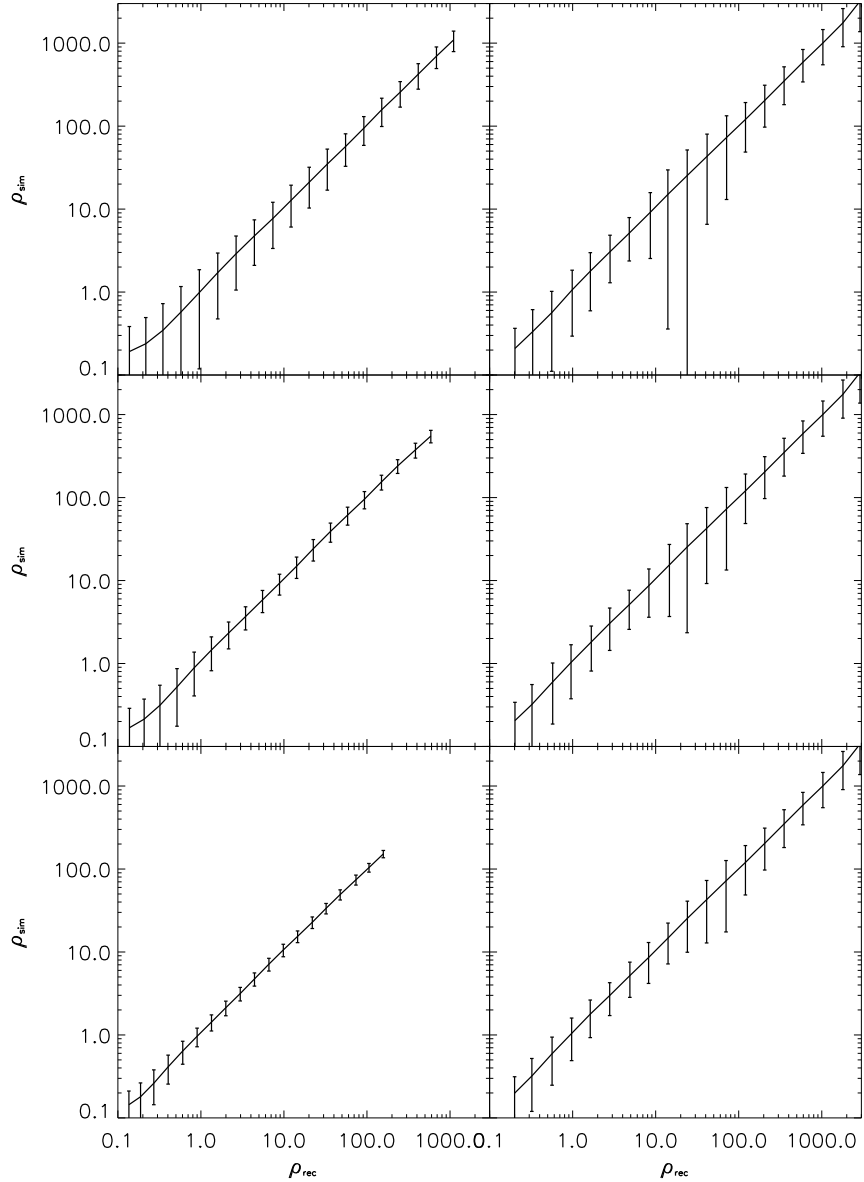} \caption{The comparison of density
between the simulation and the reconstruction. The reconstruction
here is obtained by using halo population with $M_{\rm
th}=1.68\times10^{11}\msun$ and density profiles shown in Fig.
\ref{fig_s11}. In the three left panels, the density field is
smoothed on a fixed length $1\mpc$, $2\mpc$ and $4\mpc$. In the
three right panels, the density field is smoothed with an adaptive
length $l_{\rm ad}(M_{\rm th}/2)$, $l_{\rm ad}(M_{\rm th})$ and
$l_{\rm ad}(2M_{\rm th})$ (see text for the definitions of the
smoothing length).}\label{fig_s11_dc}
\end{figure}

\begin{figure}
\epsscale{1}\plotone{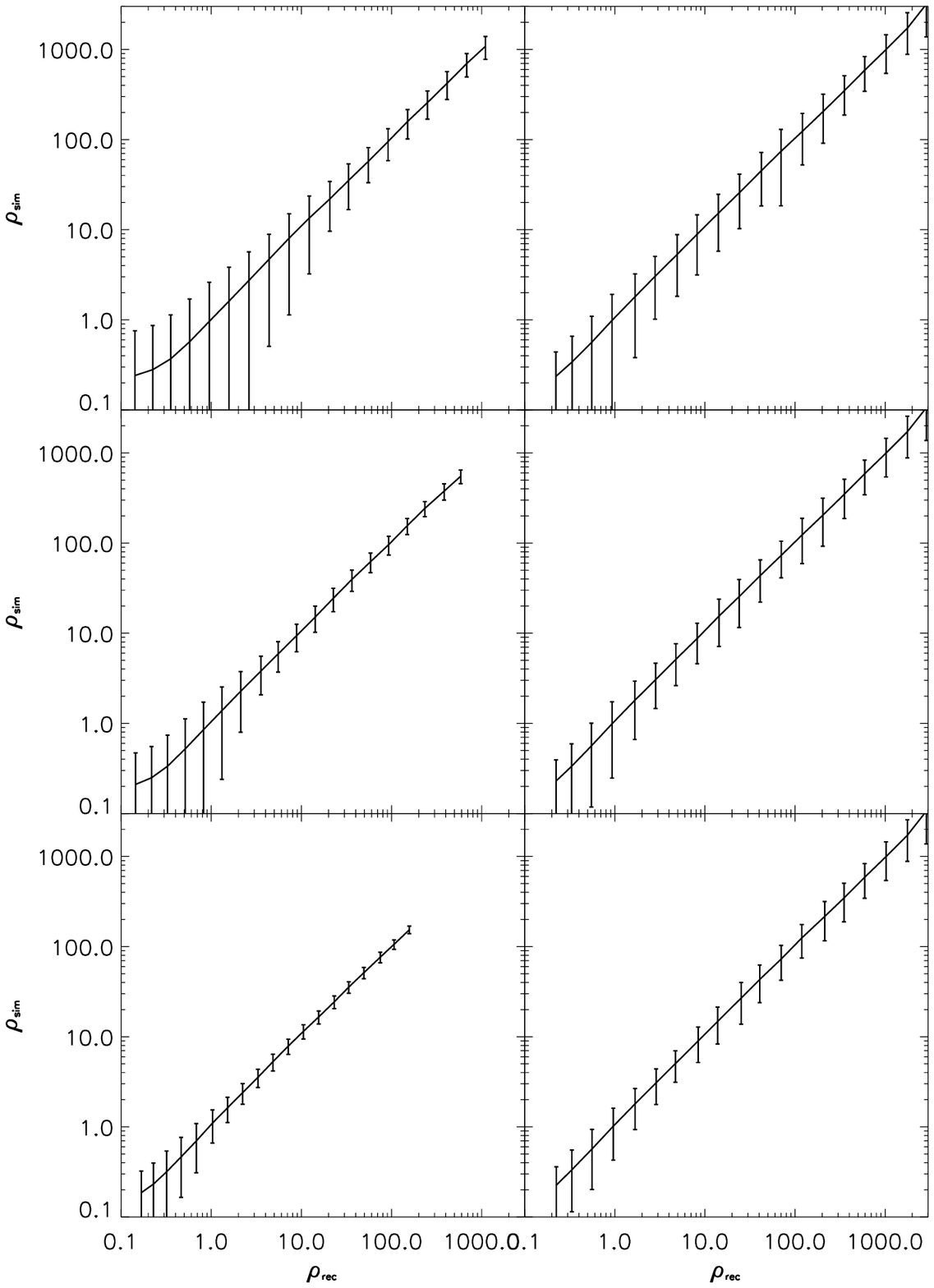} \caption{The same as Fig.
\ref{fig_s11_dc} except $M_{\rm
th}=10^{12}\msun$.}\label{fig_s12_dc}
\end{figure}

\begin{figure}
\epsscale{1}\plotone{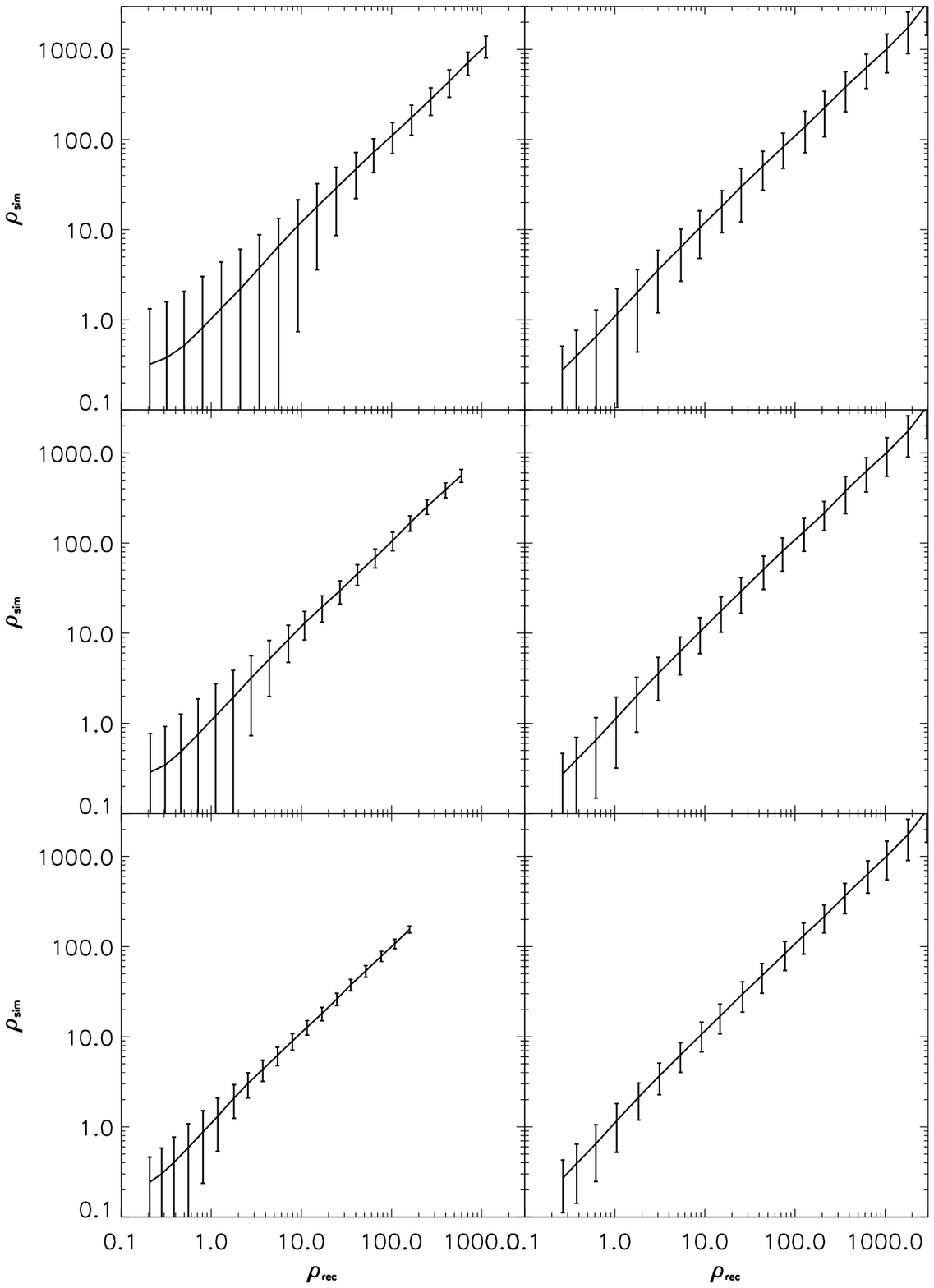} \caption{The same as Fig.
\ref{fig_s11_dc} except $M_{\rm
th}=10^{12.5}\msun$.}\label{fig_s125_dc}
\end{figure}

\begin{figure}
\epsscale{1}\plotone{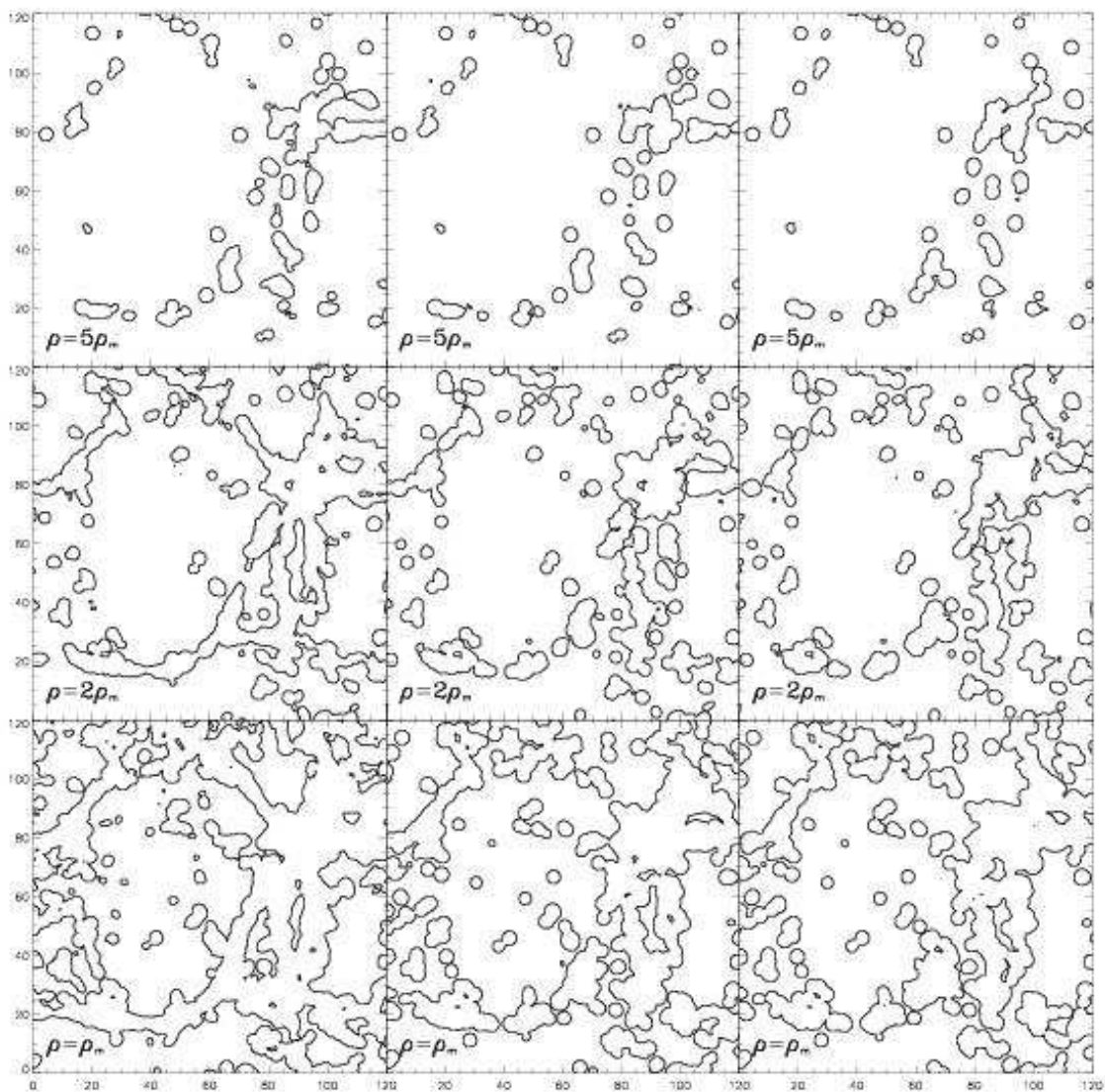} \caption{The contours at three
density levels (as indicated in the panels) of the dark matter
distribution in a slice of $120\times120\times10(\mpc)$. The left
panels show the result for simulation, the middle and right panels
are for the reconstructions based on haloes in real space and in
redshift space, respectively. The two reconstructions are made
using halo population with $M_{\rm th}=10^{12}\msun$. The contours
are based on the projections of cubes of $4\mpc$ on a side, in
which the density is above the density
threshold.}\label{fig_lsscon}
\end{figure}

\begin{figure}
\epsscale{1}\plotone{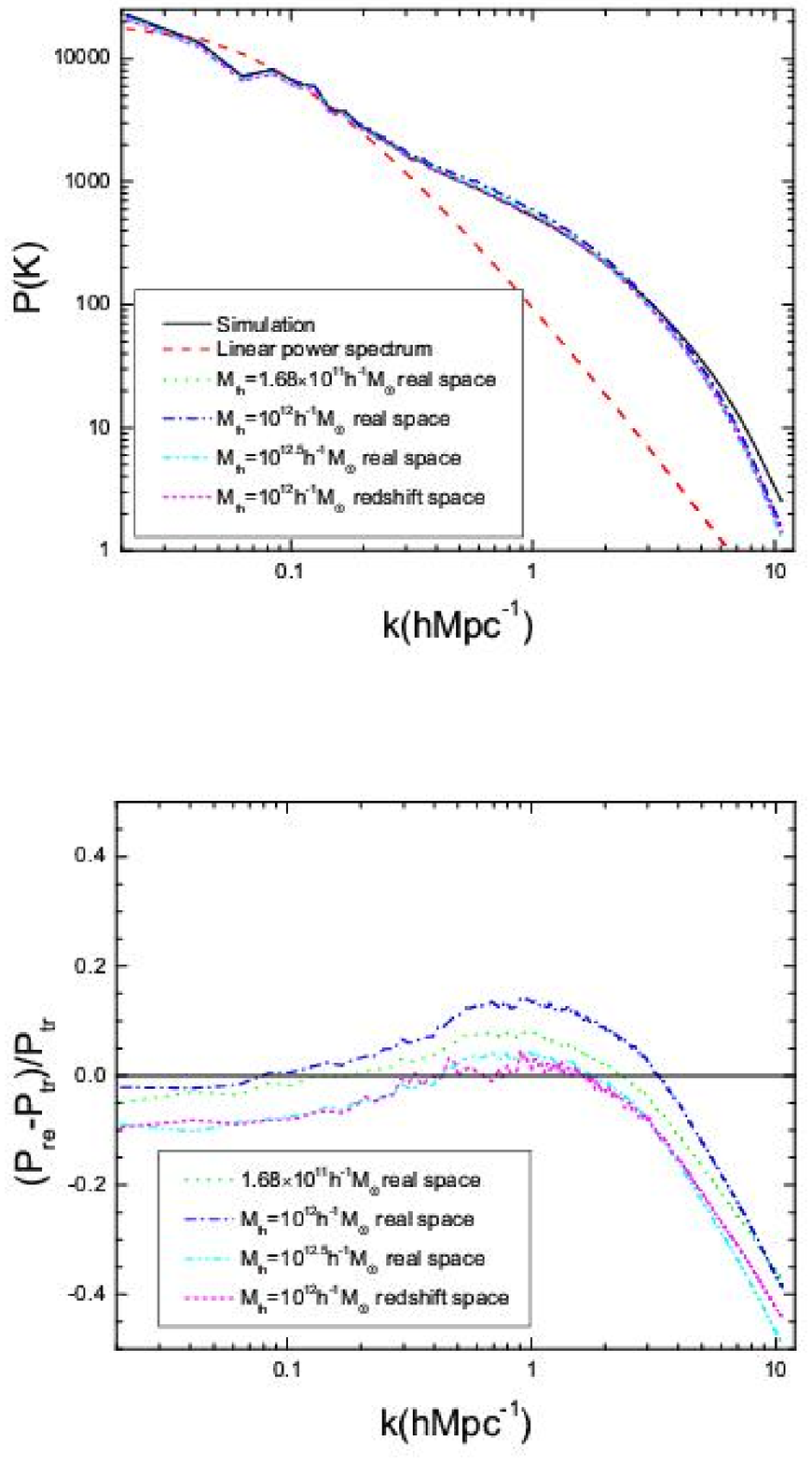} \caption{We show the true power
spectrum obtained from simulation L300, the power spectrum of the
reconstructed density field (with various halo populations, as
indicated in the panel) and the linear power spectrum in the top
panel. In the bottom panel, we show
$(\rm{P}_{re}-\rm{P}_{tr})/\rm{P}_{tr}$ as a function of
wavelength, where $\rm{P}_{re}$ and $\rm{P}_{tr}$ are the
reconstructed power spectrum and the true
 power spectrum, respectively.}\label{fig_pk}
\end{figure}

So far we have used the spherically averaged cross-correlation between
haloes and the mass in their domains to reconstruct the density field.
However,  haloes that form  in   the  cosmological density field   are
generally ellipsoidal and  should    be modelled with a   sequence  of
concentric triaxial distributions (e.g.   Jing   \& Suto 2002).    The
non-spherical properties are expected to be even more prominent in the
complementary component, where  the  mass distribution  clearly  shows
filamentary and sheet-like structure  (Fig.~\ref{fig_md}). In order to
take  such aspherical  properties  into account, we  have attempted to
model   the density  distribution   in and   around  haloes    using a
two-dimensional   profile.   A crucial  step   here  is  to define the
orientation of  the local mass distribution.   As shown in Hahn et al.
(2007b), the major axes of dark  matter haloes are strongly correlated
with the   local tidal field.    Our  own  test   using only the  mass
contained in the halo  population confirms their results.  This allows
us  to estimate two-dimensional profiles  relative  to the local tidal
axes.   We  found   that    the cross-correlation  function    becomes
increasingly  elongated up to a  scale  about 5  times the halo virial
radius, suggesting  that the  density  field around haloes  is  indeed
quite anisotropic.  We have tried  to use the two-dimensional profiles
in  our  reconstruction,  and found  that   the  improvement over  the
spherical model  is only very  modest.  There are  several reasons for
this.  First  of all, to first  order, the large-scale filamentary and
sheet-like structure has  already been taken  into account as long  as
$M_{\rm th} < M_\star$\footnote{$M_\star$  is the characteristic  mass
  scale at which the {\it RMS} of the linear density field is equal to
  $1.686$ at the present time (for our $\Lambda$CDM cosmology $M_\star
  \approx 1.0\times  10^{13} h^{-1}  M_{\odot}$)}.  This owes   to the
fact  that haloes with $M_{\rm  th}   < M_\star$ actually populate  the
filaments and sheets  that connect the more massive  haloes  with $M >
M_\star$ (see Fig.~\ref{fig_md}).  Second, the correlation between the
major axes of the local mass distribution and the local tidal field is
not   perfect, especially for low-mass   haloes, which compromises the
accuracy  of the model.   Finally,  comparing Figs.  \ref{fig_s11_dc},
\ref{fig_s12_dc}  and \ref{fig_s125_dc} one sees that  a large part of
the  discrepancy between the original field  and the reconstruction on
small scales is due to the lack of small scale structure corresponding
to haloes  with masses below $M_{\rm  th}$, which cannot  be corrected
for even with  the two-dimensional model.  Because the two-dimensional
model   is much more    complicated to  implement,   and because   the
improvement it makes to the reconstruction is not substantial, we will
only use the spherical model.

\section{Performance in redshift space}\label{sec_red}

So far we have demonstrated that our  reconstruction method works well
in real space. Unfortunately,  in  real observations the positions  of
dark haloes  represented  by   galaxy systems are  only   available in
redshift space. In this section we examine how our method performs for
a  sample of haloes (groups) in  redshift  space.  As a demonstration,
here   we focus   on the  case  with    $M_{\rm th}=10^{12}\msun$.  In
principle, we can start  with    the distribution  of haloes  in   the
redshift space, reconstruct   a density field from  this distribution,
predict the  peculiar  velocities of  haloes   using the reconstructed
density field, make  corrections  to  the  positions of  haloes,   and
iterate until a convergence is achieved. Unfortunately, this procedure
is very  time-consuming, because we have to   do the reconstruction in
each  interation. Here  we adopt  a  simplified method, using the fact
that the distribution  of dark matter haloes  can be used  to estimate
the velocity field.   It is well known  that there is bias between the
halo distribution   and   the cosmic mass  distribution,    and so the
predicted velocity field based  on the density distribution of haloes,
$\textbf{v}_h$,   is biased with respect   to the real velocity field.
Fortunately, as shown in Colombi et al. (2007), $\textbf{v}_h$ is very
tightly correlated with and directly proportional to the real velocity
field. Using the linear model given  in equation (\ref{eq_vk}), we can
write
\begin{equation}\label{eq_vhk}
\textbf{v}_h(\textbf{k})=Hf(\Omega)
\frac{i\textbf{k}}{k^2}\delta^h_{\textbf{k}}=bHf(\Omega)
\frac{i\textbf{k}}{k^2}\delta_{\textbf{k}} =b {\bf v}({\bf k})\,,
\end{equation}
where $\delta^h_{\textbf{k}}$ is   the Fourier transform of the   mass
density   contrast represented by the  mass  contained  in dark matter
haloes, and  $b$ is a constant bias  factor.  In order to estimate the
value of $b$, we assign the  mass of haloes on  grids in real-space to
obtain  $\delta^h$ and   use  equation (\ref{eq_vhk})   to  obtain the
velocities,  ${\bf v}_h$, for  all haloes.  Comparing ${\bf v}_h$ with
${\bf  v}$ based on the original  simulation we obtain $b\sim1.56$ for
$M_{\rm th}=10^{12}\msun$.  We have tried various smoothing mass scale
(SMS) to  see if the  value of $b$ is  sensitive to the SMS.  We found
that the value of  $b$ is independent  of the  SMS.  Fig.~\ref{fig_vh}
shows ${\bf v}$  versus ${\bf v}_h/b$  for haloes of different masses,
assuming a SMS of $10^{14.75}\msun$ (this SMS turns out to be the best
choice for correcting the redshift distortion, as  we will see below).
The two velocities are  very  tightly correlated, suggesting that  the
distribution of haloes can be used to predict the velocity field quite
reliably.

From equation (\ref{eq_vhk}), one can see that the bias parameter
$b$ is actually the bias of the mass distribution in halos
more massive $M_{\rm th}$ relative to the underlying mass
distribution. Thus, we can write $\delta^h=b\delta$, so that
\begin{equation}
\xi_{\rm M,m}(r,>M_{\rm th})=b\xi_{\rm m,m}(r)\,, \label{eq_ab}
\end{equation}
where $\delta^h$ and $\delta$ are the large-scale density
perturbation represented by the mass within the halo population
and the total mass, respectively; $\xi_{\rm M,m}(r,>M_{\rm th})$
is the cross-correlation between the halo mass and the underlying
mass, and $\xi_{\rm m,m}(r)$ is the autocorrelation of mass in the
universe. It is then easy to show that, on large scale, $\xi_{\rm
M,m}(r,>M_{\rm th})$ is the sum of the halo-mass
cross-correlation, $\xi_{\rm h,m}(r,M_h)$, weighted by the halos
mass:
\begin{equation}
\xi_{\rm M,m}(r,>M_{\rm th})=\frac{\int_{M_{\rm
th}}^{\infty}\xi_{\rm h,m}(r,M_h)M_hf(M_h)\,dM_h}{\int_{M_{\rm
th}}^{\infty}M_hf(M_h)\,d M_h}\,, \label{eq_ab2}
\end{equation}
where $M_hf(M_h)\,dM_h$ is the total mass in halos with masses in
the bin $[M_h,M_h+dM_h]$. Mo \& White (1996) showed that $\xi_{\rm
h,m}(r,M_h)$ is approximately parallel to $\xi_{\rm m,m}(r)$ on
large scale:
\begin{equation}
\xi_{\rm h,m}(r,M_h)=b_{\rm h}(M_h)\xi_{\rm m,m}(r)\,,
\label{eq_bh}
\end{equation}
where $b_{\rm h}(M_h)$ is the bias parameter for halos
of mass $M_{\rm h}$  (e.g. Mo \& White 1996; Jing 1998;
Sheth, Mo \& Tormen 2001). Combining the equation \ref{eq_ab}
and \ref{eq_ab2}, we obtain,
\begin{equation}
b=\frac{\int_{M_{\rm th}}^{\infty}b_{\rm
h}(M_h)M_hf(M_h)dM_h}{\int_{M_{\rm th}}^{\infty}M_hf(M_h)dM_h}\,.
\end{equation}
Obviously, the bias $b$ depends on $M_{\rm th}$ and the
cosmology in question. In Fig. \ref{fig_bh} we show the
the value of $b$ as a function of $M_{\rm th}$ predicted
by the spherical collapse model (Mo \& White 1996) and
the elliptical collapse model (Sheth, Mo \& Tormen 2001).
For comparison, we also show the results obtained directly
from the simulation. As one can see, the model predictions
are in good agreement with the simulation results, suggesting
that the value of $b$ used in our reconstruction can also be
estimated from the analytical models.

\begin{figure}
\epsscale{1}\plotone{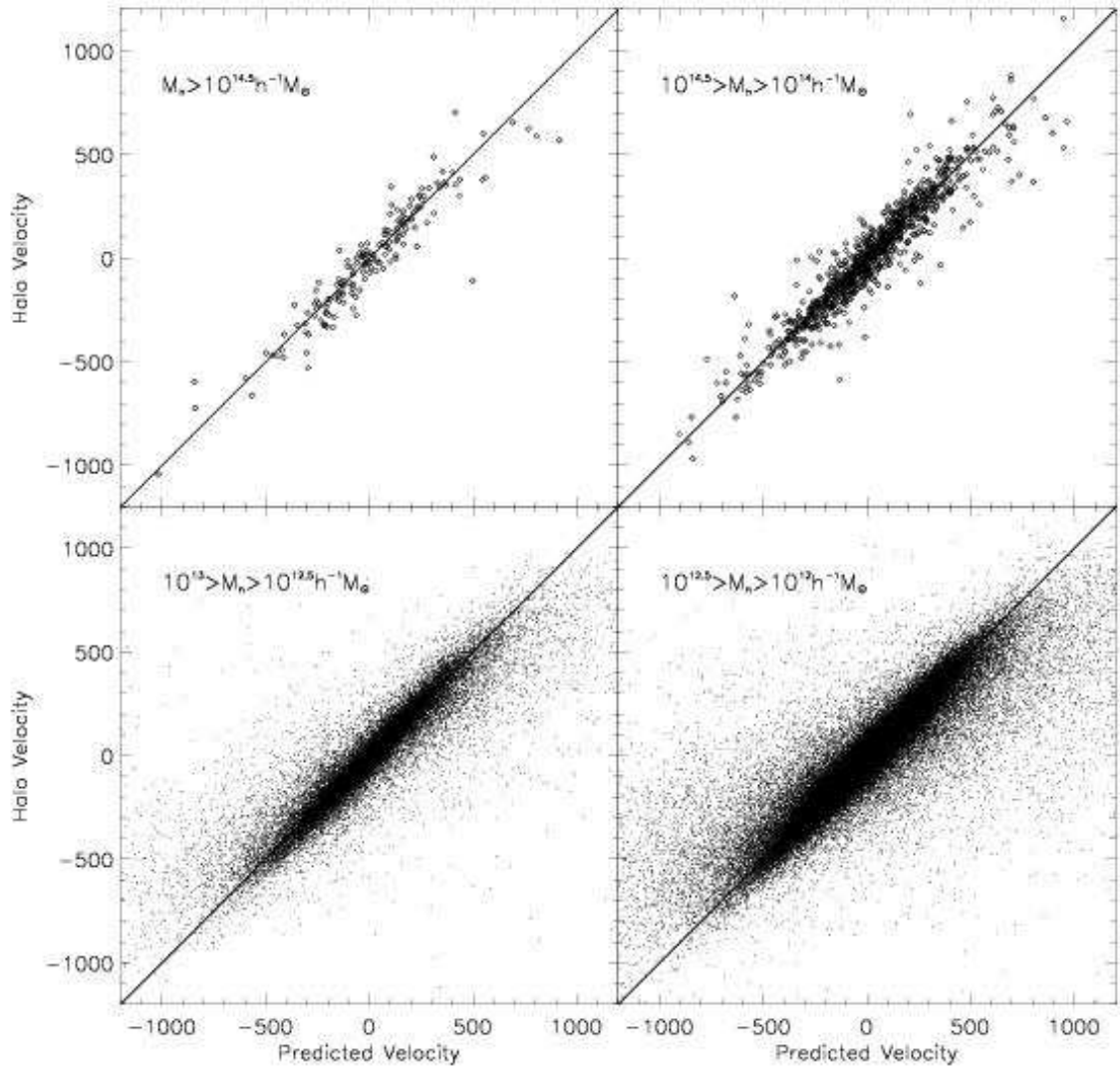}\caption{The $x$-component of the
halo velocity obtained from the simulation against the
corresponding predicted velocity by applying linear theory on the
reconstructed density field from haloes with masses above $M_{\rm
th}=1.0\times10^{12}\msun$ distributed in the real
space.}\label{fig_vel}
\end{figure}

In  the simulation, we choose the  redshift  direction to be along
the $x$-axis of the  simulation  box.  We  first use  the  halo
population in redshift space to compute the velocities,
$\textbf{v}_h$, and then use $\textbf{v}=\textbf{v}_h/b$ as
the  prediction  of the peculiar velocities  of haloes to correct
the  positions of haloes.  We iterate until convergence       is
achieved. However, as       shown   in Fig.~\ref{fig_vel}, the
velocities of haloes in general consist of two components, the
linear velocities induced  by  large scale structures and  the
non-linear velocities   induced by  small-scale  structures.
Although the former  can be easily  corrected using linear theory,
the nonlinear effect  cannot be easily  corrected, and so   it is
not very meaningful  to compare the  reconstructed field and the
original field on small scales. Because of this, we  set a minimal
SMS, $M_{\rm min}$ and use the larger of $M_h$ and $M_{\rm min}$
as the SMS for a halo of mass $M_h$. In order  to choose the value
of $M_{\rm min}$, we define the following parameter to quantify
the quality of the correction for the redshift distortion:
\begin{equation}
A_{\rm d}=\frac{\sum_{h=1}^N M_h \, \vert {\bf x}^{\rm o}_h -
{\bf x}^{\rm p}_h\vert}{\sum_{h=1}^N M_h}\,,
\end{equation}
where ${\bf x}^{\rm o}_h$ and ${\bf x}^{\rm p}_h$ are the original and
predicted positions of   halo $h$.  Thus, $A_{\rm  d}$  is the average,
mass-weighted   offset  between the    original   and predicted   halo
positions.  We found that, for the total sample of haloes, $A_{\rm d}$
is   minimized when $M_{\rm   min}$ is  about $10^{14.75}\msun$.  The
value of $A_{\rm d}$ at the minimum is about  $1.2\mpc$.  We also find
that  the  values of $M_{\rm min}$  required  for haloes  of different
masses  are quite  similar.  Notice   that  $1.2\mpc$ corresponds   to
$120{\rm   km/s}$,  which is  similar   to  the  typical error  in the
predicted  peculiar    velocities of  dark   matter haloes    shown in
Fig.~\ref{fig_vel}.    Because     of   this,    we  choose    $M_{\rm
  min}=10^{14.75}\msun$   to  estimate   the peculiar  velocities  and
correct for the redshift distortion.

With  the corrected halo  positions described  above,  we use the
same method  as   described in  Section~\ref{sec_rec}  to
reconstruct  the density field. The right three panels  of
Fig.~\ref{fig_lsscon} give a visual impression of the density
field thus obtained.  As one can see, the large  scale structure
is  well recovered. The corresponding power spectrum   is   shown
in the top panel of Fig.     \ref{fig_pk},   and   is  almost
indistinguishable from     that  obtained    from    the
real-space reconstruction.
In Fig. \ref{fig_zc} we show the reconstructed field in comparison
to the   original  density   field.    As  expected,  the
correlation between the  reconstruction  and simulation is worse
than that based on the halo population in real space, because of
the offset between the   predicted  halo positions  and  the
corresponding real positions.   However, using a    relatively
large  smoothing   length, $l_{\rm sm}\geq 5\mpc$, the scatter is
within a factor of 2. Here, we do not use the adaptive smoothing
method, because the sizes of the high-density regions, i.e. the
inner regions of the halos, are smaller than the typical offset of
the halo positions.

\begin{figure}
\epsscale{1}\plotone{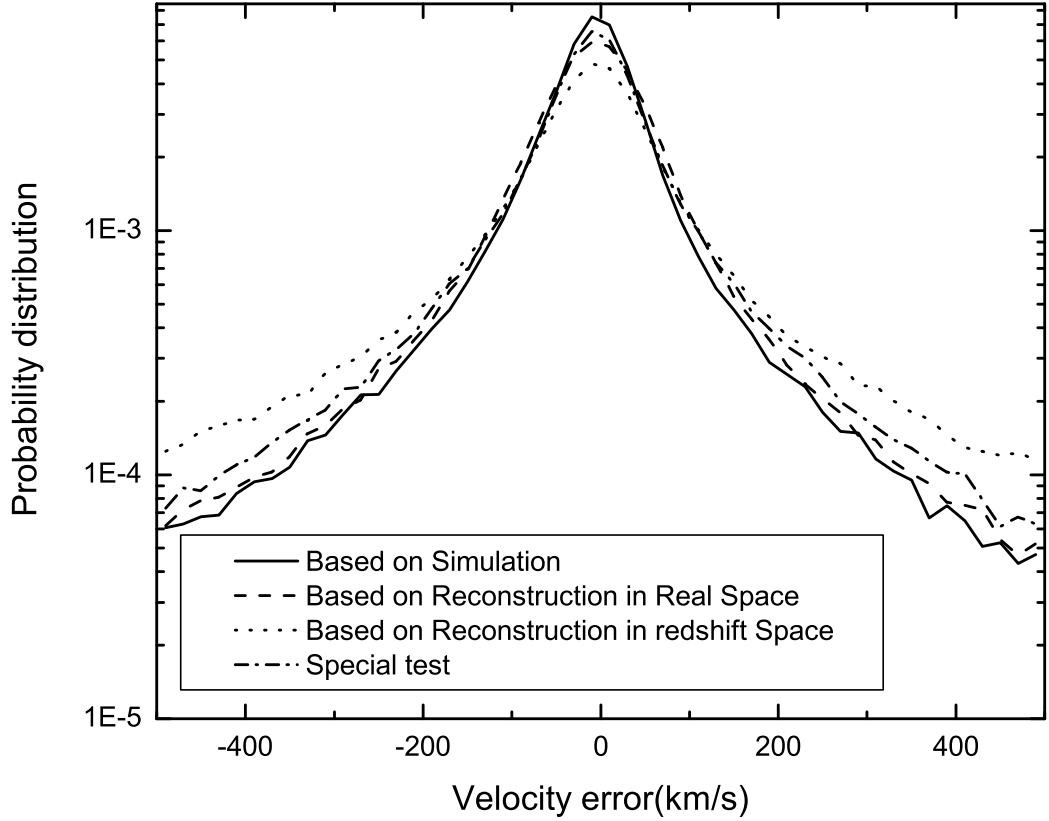} \caption{The probability
distribution of the difference between the predicted velocity and
the real velocity. The solid line represents the predicted
velocity based on the mass distribution in the original
simulation(L300). The dash and dot lines represent the predicted
velocity based on reconstructions in real space and redshift
space, respectively. Both reconstructions are made with haloes
with masses above $10^{12}\msun$. For comparison, we also show the
distribution of the velocities calculated at the real positions of
halos but using the reconstructed density field from
redshift-space data (dash-dot line; see the text for details).
}\label{fig_vd}
\end{figure}

\begin{figure}
\epsscale{1}\plotone{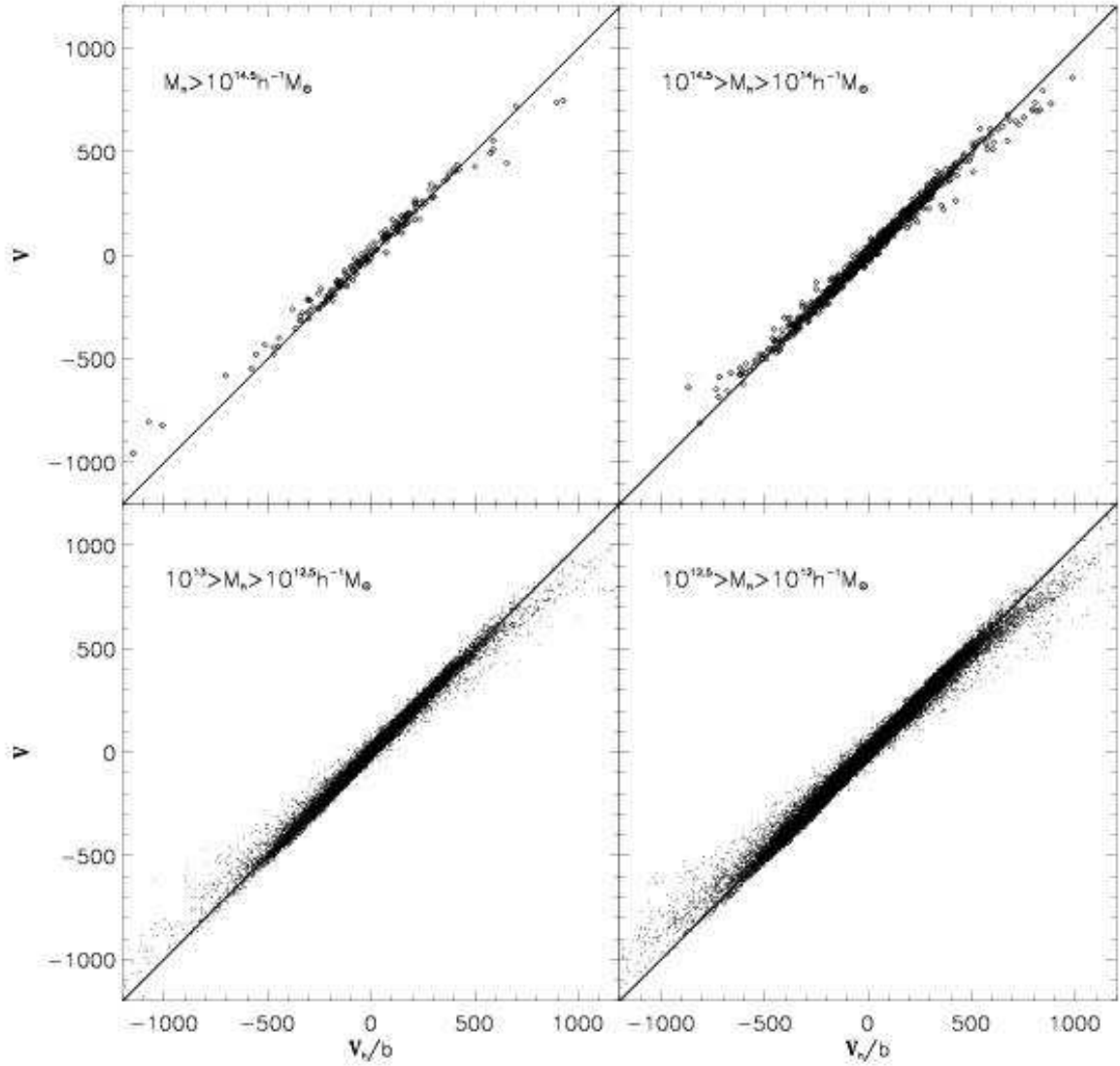} \caption{The predicted velocity,
$\textbf{v}_h$, based on halo population with masses above
$10^{12}\msun$ in real space versus $\textbf{v}$, the predicted
velocity based on the original simulated mass distribution. Here a
bias factor $b=1.56$ is used to scale $\textbf{v}_h$, and a
smoothing mass scale (SMS) of $10^{14.75}\msun$ is
used.}\label{fig_vh}
\end{figure}

\begin{figure}
\epsscale{1}\plotone{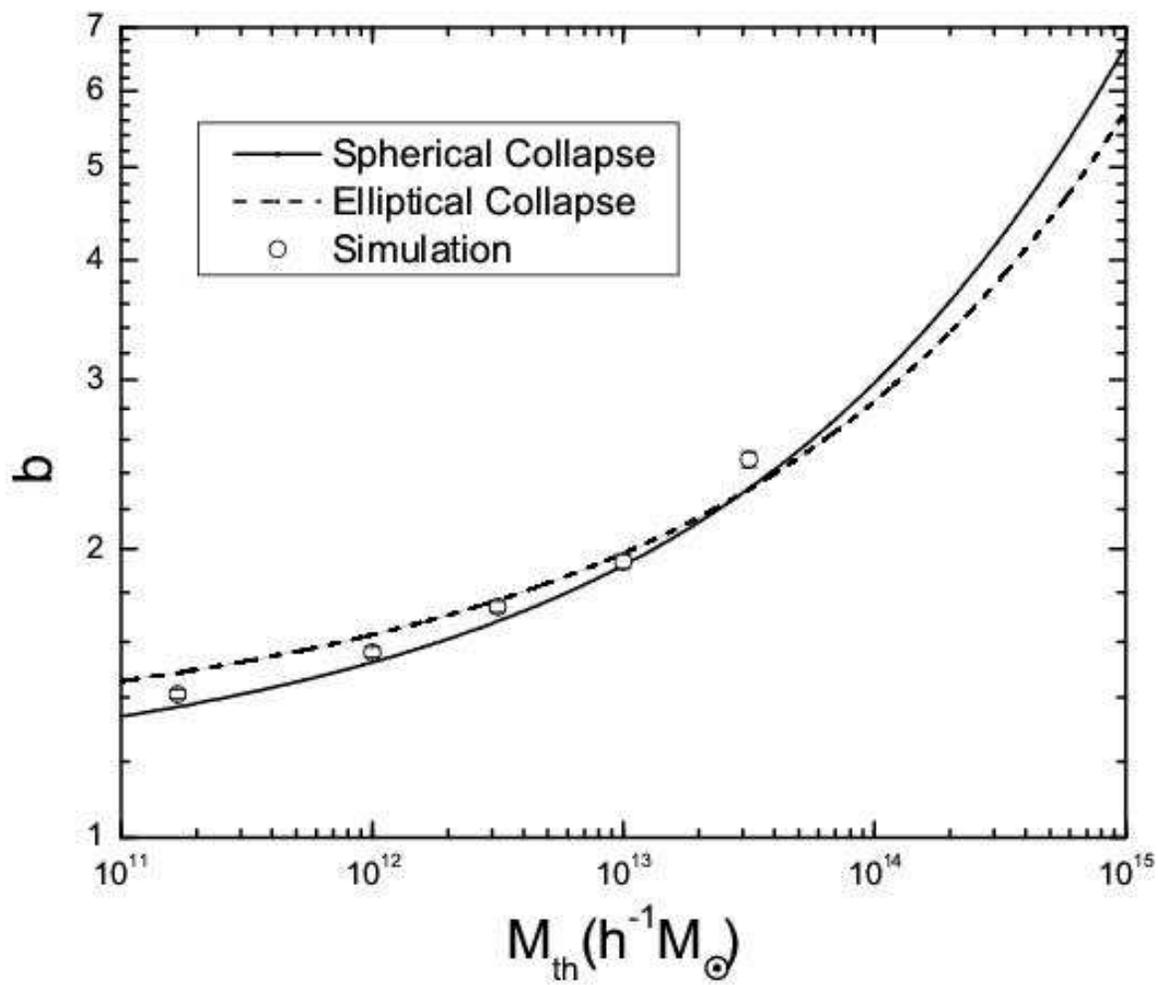} \caption{The bias parameter $b$
(defined in the text) as a function of the mass threshold $M_{\rm
th}$ for spherical collapse and elliptical collapse models
indicated in the panel. For comparison we also show the results
obtained directly from the simulation.}\label{fig_bh}
\end{figure}

\begin{figure}
\epsscale{1}\plotone{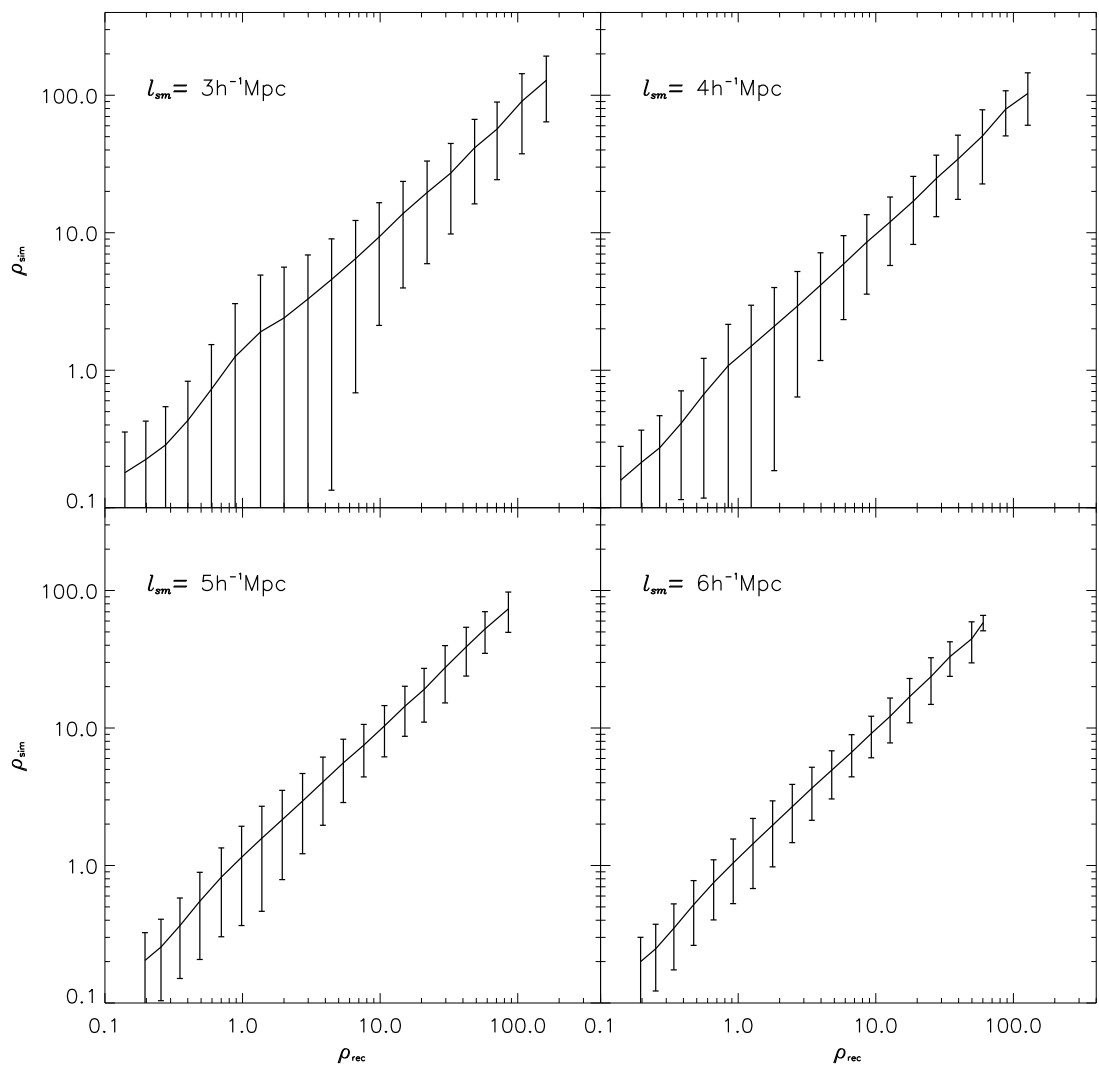} \caption{The reconstructed density
based on the halo population with $M_{\rm th}=10^{12}\msun$ in the
redshift space versus the original density in the simulation. A
fixed smoothing length is used in each panel as
indicated.}\label{fig_zc}
\end{figure}

\begin{figure}
\epsscale{1}\plotone{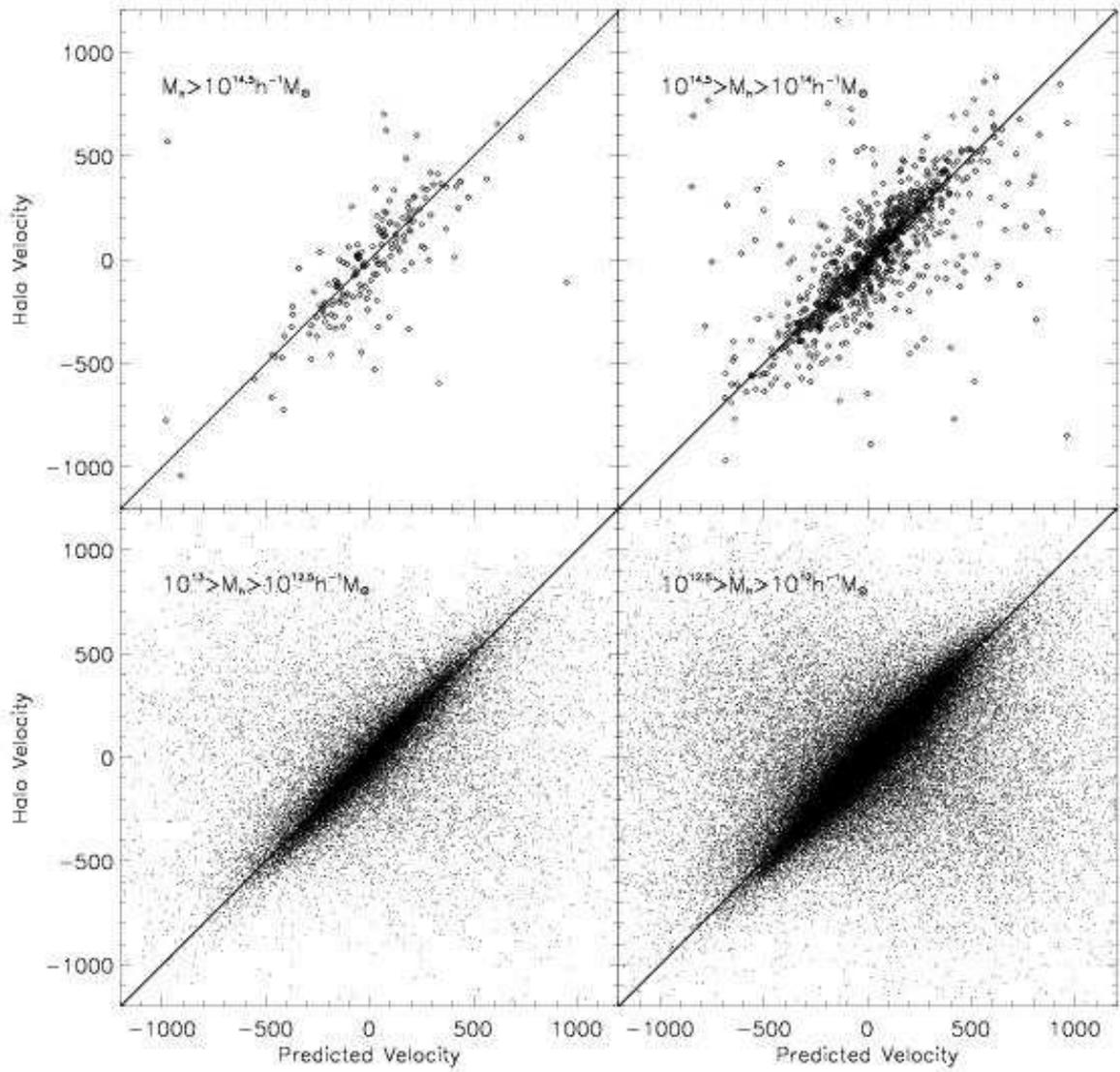} \caption{The same as Fig.
\ref{fig_vel} but the reconstruction is based on haloes
distributed in redshift space.}\label{fig_velz}
\end{figure}

We have also estimated the peculiar velocities  of haloes based on the
reconstructed field using redshift-space data. The comparison with the
true velocities   is shown in    Fig.  \ref{fig_velz}.  The  predicted
velocities  are      computed  using   the   method    described    in
Section~\ref{sec_vel}. For most  haloes, the predicted  velocities are
tightly correlated with the true velocities, but  for a small fraction
of haloes the difference is quite large.  This is also demonstrated in
Fig.  \ref{fig_vd}, where we  show the distribution  of the difference
between the predicted  and true velocities. As  one can see,  compared
with  the result obtained from the  real-space  data, the distribution
obtained with the redshift-space data has extended tails. However, the
fraction of systems in the tails  is relatively small: only about 25\%
of the systems have a velocity difference larger than $200\kms$.
There are two  kinds of effects  which can  produce such discrepancy: one
owes to the  reconstruction, the  other  is the  mismatch of the  halo
positions.  In order  to investigate which  plays a dominant  role, we
performed  the  following test.     We   use the  reconstruction    in
\textbf{redshift}-space  to calculate the  velocity field and then use
the  velocities at the  positions of the haloes in \textbf{real}-space
as the test velocities. The dot-dashed line in Fig.~\ref{fig_vd} shows
the probability distribution of  the   difference between these   test
velocities and the true halo velocities.  Note that the extended tails
have  now disappeared,  indicating that the  tails  owe mainly  to the
mismatch of the halo positions.

\begin{figure}
\epsscale{1}\plotone{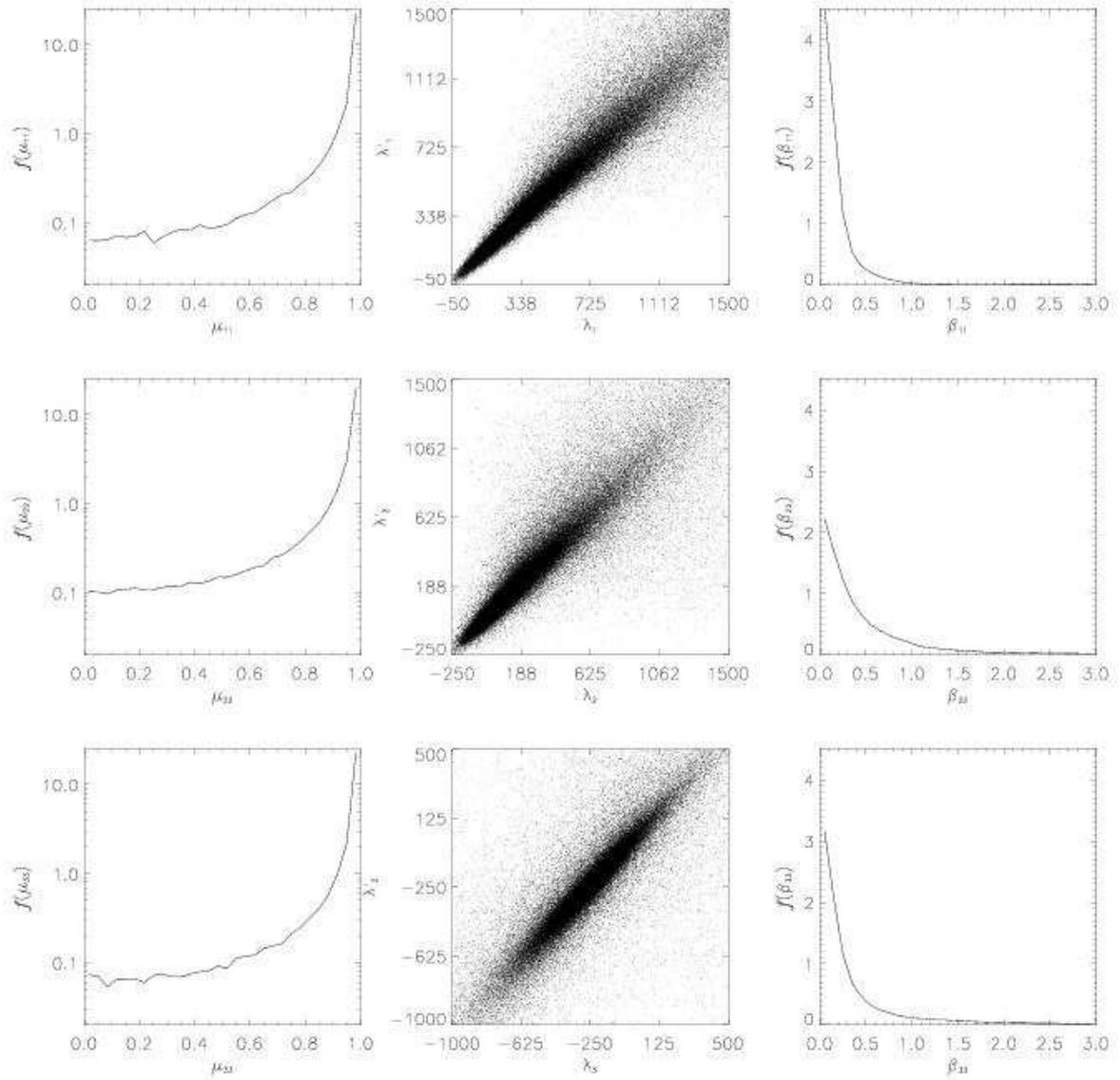} \caption{The comparison between the
tidal field obtained from the reconstructed density field (based
on haloes above $10^{12}\msun$ in redshift space) and that from
the original simulation. The three left panels show the
distribution of the dot product $\mu_i$ between $\textbf{d}'_i$
and $\textbf{d}_i$. The three middle panels show $\lambda'_i$ vs
$\lambda_i$. And the three right panels are the distribution of
$\beta_i$ (see the text for the definition).}\label{fig_tid}
\end{figure}

Finally,   we  use  the   density   field   reconstructed  from    the
redshift-space  data to compute the large  tidal field, which may play
an important role in the formation of  dark matter haloes and galaxies
(e.g. Hahn et al.  2007a,b;  Wang, Mo \&  Jing 2007). We first use CIC
scheme to generate the density field on $1024^3$ grids and then smooth
it using a Gaussian kernel.  Following Hahn et al. (2007a), we set our
SMS equal to $2 M_{\star}$. We use the Fast Fourier Transform to solve
the  Poisson equation   and apply the   second derivative  operator to
obtain  the tidal tensors.  We  compute  the eigenvalues $\lambda'_1$,
$\lambda'_2$, $\lambda'_3$ ($\lambda'_1\geq\lambda'_2\geq\lambda'_3$),
and the   corresponding eigenvectors $\textbf{d}'_1$, $\textbf{d}'_2$,
$\textbf{d}'_3$ of the tidal tensor  at the predicted position of each
halo. In order to  examine whether our  reconstruction can recover the
true tidal field, we compare the eigenvalues and eigenvectors obtained
from  the reconstructed field  based on redshift-space data with those
obtained   from  the  original simulation,   $\lambda_1$, $\lambda_2$,
$\lambda_3$   and    $\textbf{d}_1$, $\textbf{d}_2$, $\textbf{d}_3$ in
Fig.~\ref{fig_tid}. In the left three  panels we show the distribution
of  the    dot   product   $\mu_{ii}$    between  $\textbf{d}'_i$  and
$\textbf{d}_i$   ($i=1,2,3$).   As  one  can   see,  the reconstructed
$\textbf{d}'_i$s  are strongly   aligned with  the  corresponding true
$\textbf{d}_i$s.   The  values  of  $\lambda'_i$s   are also  strongly
correlated with $\lambda_i$s, as  shown in the  three middle panels of
Fig. \ref{fig_tid}. To quantify the correlation, we define a parameter
$\beta_i=|\lambda'_i-\lambda_i|/|\lambda_i|$,    and      show     the
distribution of $\beta_i$s  in the three  right panels. The values  of
$\beta_i$ are quite small  for the majority of  the haloes. Note that,
on average,  $\beta_{11} < \beta_{33}  < \beta_{22}$; this simply owes
to the fact that $|\lambda_1| > |\lambda_3| > |\lambda_2|$ on average.

\section{Discussion}
\label{sec_app}

In  this paper, we have used  $N$-body simulations  to demonstrate the
promise of using the distribution of dark matter haloes to reconstruct
the underlying density  field. This paper  serves  only as a proof  of
concept,   and the applications of the   method to real   data will be
presented  in   forthcoming papers. In   this   discussion section, we
outline some  of   the  questions  that    can be  addressed  with   a
well-reconstructed cosmic density field.

The largest redshift sample of galaxies now available is that
given by the SDSS.  With the SDSS group (halo) catalogue
constructed by Yang et al.  (2007),   it is   in  principle
straightforward to  apply   the reconstruction   procedures
described  above to   obtain  the current density  field in the
SDSS  volume.  Since  the observed positions of haloes  are in
redshift space, one  has to take  into account redshift
distortion. As discussed  above, the structures  traced by dark
matter haloes are  expected to be in the  mildly non-linear
regime, and so a correction for peculiar velocities can be made
using the reconstructed mass  distribution  in  an   iterative
way.  In  addition to redshift distortion, the  observational
group  sample   also suffers    from incompleteness   and
contaminations  due  to   observational selection effects  and
errors introduced  by  the group  finder,  from  the uncertainties
in the  halo mass assignments to  groups, and from the boundary
effects. The results of several tests demonstrated that the
method of halo-mass assignment is statistically reliable.
Yang et al. (2007) tested the group finder using mock catalogs and
found that the group finder successfully selects more than 90 percent
of all the true halos in N-body simulations, and that
the scatter between the estimated and true halo masses is
about 0.3 dex.  Yang et al. (2005) and Wang et al. (2008)
estimated the clustering bias of halos as a function of group mass,
and found very good agreement with the mass dependence of halos
in $N$-body simulation, which is not expected if the halo masses
assigned to galaxy groups are in serious error.
Finally, in a recent investigation, Li et al. (2008) found that
the group catalog selected from the SDSS, combined  with the masses
assigned to groups can reproduce the galaxy-galaxy weak lensing
signal observed by Mandelbaum et al. (2006). Nevertheless,
the impact of all  the factors should be tested and  quantified with
the  use of detailed  mock catalogues, such as  the ones presented
in Yang et al. (2005; 2007). We will come back to this in a
forthcoming paper.

Recent results   from  $N$-body simulations  show  that  the formation
history of a  dark matter halo   can be significantly  affected by its
large-scale environment (Gao  et al. 2005; Wechsler  et al. 2006; Wang
et al.  2007; Jing, Suto  \& Mo 2007).  Furthermore, the structure and
kinematics of dark   matter haloes,  such   as its  shape and  angular
momentum, may also be correlated with  the large-scale environments of
haloes (e.g.  Faltenbacher et al. 2007). Since  galaxies form  in dark
matter haloes, the  environmental dependence   of halo properties   is
expected to  produce observable correlations between galaxy properties
and  large-scale  structure. Indeed,  recent gas-dynamical simulations
demonstrate that the accretion of cold gas  into dark matter haloes to
form galaxies is dominated by flows along filaments (Kere{\v s} et al.
2005). With our reconstructed density field, we will be able to define
in detail  the environments of all  the galaxies in  the SDSS based on
their neighboring galaxies  as well as  on  the tidal  fields owing to
large scale structure. It is  then straightforward to analyze how  the
properties of individual galaxies, such as shape, orientation, angular
momentum,  size, and gas content,  are correlated with the large-scale
environments. Some investigations  along this  line have been  carried
out successfully using different  redshifts surveys and methods  (e.g.
Lee \& Erdo{\u g}du 2007; Longo 2007; Lee \& Lee 2008).

A key  step in  understanding  galaxy formation in  the cosmic density
field is to study the distribution,  state and chemical composition of
the   diffuse  gas, i.e. gas   that  has  not  been  incorporated into
galaxies, and its  relationship to the  galaxy population.  There have
been many observational programs dedicated  to various aspects of this
IGM component. Extensive   X-ray observations have been   conducted to
study the hot gas associated with clusters and rich groups of galaxies
but one expects  the  total mass associated   with such systems  to be
small.  As shown in Mo \& White (2002), at the present time only $\sim
10\%$  of the  cosmic   mass  is in   virialized  haloes with   virial
temperatures  above 1  KeV.   About $70\%$   of all  the  mass  is  in
virialized haloes with virial temperatures  below $10^{6}$ K, too cold
to be studied using X-ray observations. A  more promising, and perhaps
the only    way, to study the   bulk  of the  diffuse IGM   is through
absorption lines.  The   capability    of  this  approach  has    been
demonstrated very convincingly at high redshift, where observations of
Lyman-$\alpha$ and  metal absorption lines using  optical spectroscopy
have identified almost all of the baryonic component and have provided
important clues about the  nature of the IGM at  $z\sim 3$ (e.g. Rauch
et al. 1998 and references therein).  At low redshift ($z\sim 0$), the
situation  is more complicated.  First of  all,  one needs UV or X-ray
spectroscopy, meaning space-based   observations, to study the  common
atomic absorption lines.  Secondly, at   low-$z$ a larger fraction  of
the  IGM   may have been  affected    by gravitational   collapse  and
non-gravitational  processes,     such as  star     formation and  AGN
activities.  The  structure, state and   composition of the gas to  be
studied is, therefore,   more  complex making the   interpretation  of
observational results more challenging. Observations with the {\it Far
  Ultraviolet Spectroscopic Explorer} ({\it FUSE}) and the {\it Hubble
  Space  Telescope} ({\it HST})  and  its {\it Space Telescope Imaging
  Spectrograph} ({\it STIS}) have demonstrated the promise of using UV
absorption systems to identify the diffuse component of the IGM at
low redshifts.  These  observations have so  far revealed  a wide
array of absorptions lines, ranging from low ions such as  HI all
the way up to highly ionized species, such  as  OVI and  NeVIII
(e.g. Stocke e  al. 2004;  Tripp  \&  Bowen  2005   and references
therein),  presumably associated with the warm-hot medium  seen in
gas-dynamical simulations (e.g. Cen  \& Ostriker 1999;  Dav\'e et
al. 2001).  With spectrographs aboard the {\it Chandra} and {\it
XMM/Newton}  X-ray telescopes, it is now also possible   to  probe
the IGM  through   very  highly-ionized species, such as OVII,
OVIII,  and NeIX.  The  current sample is still too small to allow
a systematic  and detailed exploration of the IGM, but fortunately
the  situation is expected to improve  soon with the installation
of the {\it  Cosmic Origins Spectrograph} ({\it COS}) on the {\it
HST} and the launch of a future X-ray satellite, {\it X-ray
Evolving-Universe Spectroscopy}(Kawahara et al. 2006); the sample
of the absorption systems should increase by an order-of-magnitude
or more. Clearly, it is important to examine in detail  how such
observations can be used to  identify the missing component of the
IGM  at low-$z$ and to study its properties  in a systematic way.

Absorption   studies at  low-redshift have  suffered   from  a lack of
suitable  background sources, i.e. quasars   and gamma-ray bursts, and
from the difficulty in obtaining high-quality UV spectroscopy.  It is,
therefore, imperative to have as much  theoretical and empirical input
as possible both to design  an optimal  observational strategy and  to
help interpret  the limited   amount of  observational data.   With  a
detailed    map of   the   local  density   field   obtained  from the
reconstruction  method     described here, we    can  obtain  detailed
information  about  the environment  in  which the  absorption occurs.
This is particularly useful for interpreting the observational data of
low-$z$ absorption systems,  because   here the  sample  is small  and
cosmic variance is a major concern. The uncertainties are minimized if
comparisons   between the observations and   the model predictions are
made for systems with the same large-scale environments.

\section{Summary}\label{sec_sum}

In this paper we have developed a method to reconstruct cosmic
density field from the distribution of dark mater haloes  above a
certain mass threshold,  $\rm M_{\rm th}$. Our     method uses the
fact that   the statistical  properties    of the   large-scale
structure   are  well represented  by the current  $\Lambda$CDM
model. In order to describe the distribution of the  mass in and
around  dark matter  haloes, each volume element in space is
assigned to the domain  of the nearest halo according to a
distance measure that is scaled by the virial radius of the halo.
The density profiles associated with  dark matter haloes are then
modelled using the cross-correlation function between dark matter
haloes  and the mass distribution  within their domains. In this
paper of proof-of-concept,   we use   two  sets of high-resolution
N-body simulation to calculate such profiles. Within  the virial
radii, these density profiles  are well represented by  the   NFW
profile, and  on larger scales  the profiles are  comparable to
those obtained by Prada et al. (2006). These  density profiles are
used  to  sample  the mass distribution in the  domains of all
haloes  to reconstruct the density field that   are not associated
with the   virialized haloes above the mass threshold. Since our
reconstruction uses the distribution of dark matter haloes, it
ensures that the reconstructed density field is the same  as that
traced  by the halo  population, and on  small scales it
reproduces  the  NFW profiles within  individual   haloes.  We
have considered   three cases, using  $M_{\rm
th}=1.68\times10^{11}$, $10^{12}$ and $10^{12.5}\msun$,
respectively.  The  later two  values represent the lower-mass
limit  of complete group  samples that can be selected from SDSS
to a redshift of $z \sim0.1$ and 0.15. Clearly, the density field
is better reconstructed with a smaller $M_{\rm th}$. A comparison
between the reconstructed  field  and the original field shows
that our method can reconstruct the density accurately, with an
error typically of  $50\%$, on mass scales comparable to or larger
than $(1\to 2)M_{\rm th}$. The  reconstructed density field
can be  used to estimate the peculiar
velocities of dark matter haloes with a typical error of $\sim 100{\rm
  km\,s^{-1}}$.

We also test the reliability of  our method working in redshift space.
To do this, we start with the positions of haloes in redshift space to
estimate the peculiar velocities and  use iterations until convergence
is achieved.  The positions of haloes in  the final iteration are then
used to do the reconstruction.  Because  of the error in the predicted
peculiar velocities, there are offsets between the predicted positions
of haloes  and the real positions. This  leads to mismatch between the
reconstructed  field and the original  field on small scales. However,
the   density and   velocity field on   large scales   are still  well
reconstructed.

We have  outlined  some potential  applications  of our reconstruction
method. Our final goal is to apply this  method to observational data,
such  as the SDSS  group  catalogue. With a well-reconstructed density
field in the local universe, we can study in detail how the properties
of  galaxies  are  affected by   their large-scale  environments.  Our
reconstruction  can also be used to  study the correlation between the
gas component to be revealed by QSO absorption systems at low-redshift
with the large-scale environments.

\section*{Acknowledgments}

We acknowledge the support from  the  Supercomputing Center of
USTC. HYW would like to acknowledge the support  of the Knowledge
Innovation Program  of the Chinese Academy of Sciences, Grant No.
KJCX2-YW-T05 and NSFC  10643004. HJM would like to acknowledge the
support of NSF AST-0607535, NASA AISR-126270 and NSF IIS-0611948.
YPJ is supported by NSFC (10533030), by the Knowledge Innovation
Program of CAS (No. KJCX2-YW-T05), and by 973 Program (N
o.2007CB815402).

\appendix{\textbf{Appendix A}: Note on the effective volume}

Delaunay Density Estimator Method was used by Schaap \& van de
Weygaert (2000, hereafter SW2000; see also Wang's phd thesis)
to construct a continuous density field based on a sample of
discrete data points and has been demonstrated to have optimal
performance in both high and low density regions than the
conventional methods, such as the grid-based TSC method and
the SPH smoothing kernel method. It \emph{divides the space into a
unique and volume-covering network of mutually disjunct Dealunay
tetrehedra} (SW2000, see their figure 1) and the vertices of the
tetrahedra are just the CDM particles. So the density at a vertex,
e.g. at the position of particle `p', is
 \begin{equation}\label{eq_rhop}
 \rho_p=\frac{4M_p}{W_{Vor,p}}\,.
 \end{equation}
According to this formula we define the effective volume of
particle `p' as
\begin{equation}\label{eq_vp}
V_p=\frac{W_{Vor,p}}{4}\,,
\end{equation}
where $W_{Vor,p}$ is the sum of volume of all tetrahedra
associated with this vertex. Obviously, the value of $W_{Vor,p}$
is determined by the nearby particles distribution.
The method is fully adaptive and follows the mass conservation
and volume conservation (see SW2000 for more details).

 We use the publicly released software Qhull
\footnote{downloaded from http://www.qhull.org/}
(Barber et al. 1996) to generate the Delaunay tetrahedra and
use equation (\ref{eq_vp}) to calculate the effective volume
associated with each CDM particle.

\appendix{\textbf{Appendix B}: Note on the SPH smoothing method}

Since the density gradient is very steep within the virial radius
of halos, the density at the center of the `base-grid', with a size
of $250\kpc$ on a side, cannot be used to represent the
average density of the whole `base-grid'. In order to recover the
density properly, we divide the `base-grid' into $7^3$ sub-grids,
with the grid-size comparable to the softening length used in the
simulation. Then we use SPH method to smoothing the particle
distribution on the sub grid. The smoothing is performed using the
$N_{\rm sph}=32$ nearest neighbors to each sub-grid-point. We
adopt the following SPH kernel (Monaghan 1992),
 \[ \zeta(x)=\left\{
           \begin{array}{ll}
               1-1.5x+0.75x^3\,,  (0\leq x\leq1) \\
               0.25(2-x)^2\,,  (1\leq x\leq2) \\
               0   \,,(x>2)
            \end{array}
            \right.\,,
 \]
and obtain the density $\rho$ on a sub grid site,
$\textbf{r}_{\rm g}$, using
\begin{equation}
\rho(\textbf{r}_{\rm g})=\frac{8M_p}{\pi R^3_{\rm
sph}(\textbf{r}_{\rm g})}\sum_{i=1}^{N_{\rm
sph}}\zeta\left[\frac{r_i}{R_{\rm sph}(\textbf{r}_{\rm
g})}\right]\,,
\end{equation}
where $R_{\rm sph}(\textbf{r}_{\rm g})$ is half the distance from
${\bf r}_g$ to the farthest neighbor of the sub grid among the
$N_{\rm sph}$ particles, and $r_i$ is the distance of the $i$th
particle to the sub grid site. Finally we average over the
densities of all the $7^3$ sub-grids to obtain the density of the
corresponding `base-grid'.

\newpage

\label{lastpage}


\begin{thebibliography}{}

\bibitem[\protect\citeauthoryear{Barber et
al.}{1996}]{ACM Trans. on Mathematical Software,
22(4),469--483}Barber, C. B., Dobkin, D. P., \& Huhdanpaa, H. T.
(1996). The Quickhull algorithm for convex hulls. ACM Trans. on
Mathematical Software, 22(4),469--483

\bibitem[\protect\citeauthoryear{Bardeen et
al.}{1986}]{1986ApJ...304...15B} Bardeen J.~M., Bond J.~R., Kaiser
N., Szalay A.~S., 1986, ApJ, 304, 15

\bibitem[\protect\citeauthoryear{Bartelmann
\& Schneider}{2001}]{2001PhR...340..291B} Bartelmann M., Schneider
P., 2001, PhR, 340, 291

\bibitem[\protect\citeauthoryear{Berlind
\& Weinberg}{2002}]{2002ApJ...575..587B} Berlind A.~A., Weinberg
D.~H., 2002, ApJ, 575, 587

\bibitem[\protect\citeauthoryear{Bullock et
al.}{2001}]{2001MNRAS.321..559B} Bullock J.~S., Kolatt T.~S.,
Sigad Y., Somerville R.~S., Kravtsov A.~V., Klypin A.~A., Primack
J.~R., Dekel A., 2001, MNRAS, 321, 559

\bibitem[\protect\citeauthoryear{Cen \&
Ostriker}{1999}]{1999ApJ...514....1C} Cen R., Ostriker J.~P.,
1999, ApJ, 514, 1

\bibitem[\protect\citeauthoryear{Croton et al.}{2006}]{2006MNRAS.365...11C}
Croton D.~J., et al., 2006, MNRAS, 365, 11

\bibitem[\protect\citeauthoryear{Cole et al.}{2000}]{2000MNRAS.319..168C}
Cole S., Lacey C.~G., Baugh C.~M., Frenk C.~S., 2000, MNRAS, 319,
168

\bibitem[\protect\citeauthoryear{Colless et
al.}{2001}]{2001MNRAS.328.1039C} Colless M., et al., 2001, MNRAS,
328, 1039

\bibitem[\protect\citeauthoryear{Colombi, Chodorowski, \&
Teyssier}{2007}]{2007MNRAS.375..348C} Colombi S., Chodorowski
M.~J., Teyssier R., 2007, MNRAS, 375, 348

\bibitem[\protect\citeauthoryear{Cooray}{2006}]{2006MNRAS.365..842C} Cooray
A., 2006, MNRAS, 365, 842

\bibitem[\protect\citeauthoryear{Cooray \&
Sheth}{2002}]{2002PhR...372....1C} Cooray A., Sheth R., 2002, PhR,
372, 1

\bibitem[\protect\citeauthoryear{Dav{\'e} et
al.}{2001}]{2001ApJ...552..473D} Dav{\'e} R., et al., 2001, ApJ,
552, 473

\bibitem[\protect\citeauthoryear{Erdo{\u g}du et
al.}{2004}]{2004MNRAS.352..939E} Erdo{\u g}du P., et al., 2004,
MNRAS, 352, 939

\bibitem[\protect\citeauthoryear{Evrard, Metzler,
\& Navarro}{1996}]{1996ApJ...469..494E} Evrard A.~E., Metzler
C.~A., Navarro J.~F., 1996, ApJ, 469, 494

\bibitem[\protect\citeauthoryear{Faltenbacher et
al.}{2007}]{2007ApJ...662L..71F} Faltenbacher A., Li C., Mao S.,
van den Bosch F.~C., Yang X., Jing Y.~P., Pasquali A., Mo H.~J.,
2007, ApJ, 662, L71

\bibitem[\protect\citeauthoryear{Fisher et al.}{1995}]{1995MNRAS.272..885F}
Fisher K.~B., Lahav O., Hoffman Y., Lynden-Bell D., Zaroubi S.,
1995, MNRAS, 272, 885

\bibitem[\protect\citeauthoryear{Gao, Springel, \&
White}{2005}]{2005MNRAS.363L..66G} Gao L., Springel V., White
S.~D.~M., 2005, MNRAS, 363, L66

\bibitem[\protect\citeauthoryear{Gastaldello et
al.}{2007}]{2007ApJ...669..158G} Gastaldello F., Buote D.~A.,
Humphrey P.~J., Zappacosta L., Bullock J.~S., Brighenti F.,
Mathews W.~G., 2007, ApJ, 669, 158

\bibitem[\protect\citeauthoryear{Hahn et al.}{2007}]{2007MNRAS.375..489H}
Hahn O., Porciani C., Carollo C.~M., Dekel A., 2007a, MNRAS, 375,
489

\bibitem[\protect\citeauthoryear{Hahn et al.}{2007}]{2007MNRAS.381...41H}
Hahn O., Carollo C.~M., Porciani C., Dekel A., 2007b, MNRAS, 381,
41

\bibitem[\protect\citeauthoryear{Hockney
\& Eastwood}{1981}]{1981csup.book.....H} Hockney R.~W., Eastwood
J.~W., 1981, csup.book,


\bibitem[\protect\citeauthoryear{Jing}{1998}]{1998ApJ...503L...9J} Jing
Y.~P., 1998, ApJ, 503, L9

\bibitem[\protect\citeauthoryear{Jing, Mo, \&
Boerner}{1998}]{1998ApJ...494....1J} Jing Y.~P., Mo H.~J., Boerner
G., 1998, ApJ, 494, 1

\bibitem[\protect\citeauthoryear{Jing \& Suto}{2002}]{2002ApJ...574..538J}
Jing Y.~P., Suto Y., 2002, ApJ, 574, 538

\bibitem[\protect\citeauthoryear{Jing, Suto,
\& Mo}{2007}]{2007ApJ...657..664J} Jing Y.~P., Suto Y., Mo H.~J.,
2007, ApJ, 657, 664

\bibitem[\protect\citeauthoryear{Kang et al.}{2005}]{2005ApJ...631...21K}
Kang X., Jing Y.~P., Mo H.~J., B{\"o}rner G., 2005, ApJ, 631, 21

\bibitem[\protect\citeauthoryear{Kauffmann, White,
\& Guiderdoni}{1993}]{1993MNRAS.264..201K} Kauffmann G., White
S.~D.~M., Guiderdoni B., 1993, MNRAS, 264, 201

\bibitem[\protect\citeauthoryear{Kauffmann et
al.}{2004}]{2004MNRAS.353..713K} Kauffmann G., White S.~D.~M.,
Heckman T.~M., M{\'e}nard B., Brinchmann J., Charlot S., Tremonti
C., Brinkmann J., 2004, MNRAS, 353, 713

\bibitem[\protect\citeauthoryear{Katz, Weinberg,
\& Hernquist}{1996}]{1996ApJS..105...19K} Katz N., Weinberg D.~H.,
Hernquist L., 1996, ApJS, 105, 19

\bibitem[\protect\citeauthoryear{Kawahara et
al.}{2006}]{2006PASJ...58..657K} Kawahara H., Yoshikawa K., Sasaki
S., Suto Y., Kawai N., Mitsuda K., Ohashi T., Yamasaki N.~Y.,
2006, PASJ, 58, 657

\bibitem[\protect\citeauthoryear{Kere{\v s} et
al.}{2005}]{2005MNRAS.363....2K} Kere{\v s} D., Katz N., Weinberg
D.~H., Dav{\'e} R., 2005, MNRAS, 363, 2

\bibitem[\protect\citeauthoryear{Lahav et al.}{1991}]{1991MNRAS.251..128L}
Lahav O., Lilje P.~B., Primack J.~R., Rees M.~J., 1991, MNRAS,
251, 128

\bibitem[\protect\citeauthoryear{Lavaux}{2008}]{2008arXiv0801.4208L} Lavaux
G., 2008, arXiv, 801, arXiv:0801.4208

\bibitem[\protect\citeauthoryear{Lee
\& Erdogdu}{2007}]{2007ApJ...671.1248L} Lee J., Erdogdu P., 2007,
ApJ, 671, 1248

\bibitem[\protect\citeauthoryear{Lee
\& Lee}{2008}]{2008arXiv0801.1558L} Lee J., Lee B., 2008,
arXiv, 801, arXiv:0801.1558

\bibitem[\protect\citeauthoryear{Li et al}{2008}]{2008MNRAS...submitted}
Li, R., Mo, H.J., Fan, Z., Cacciato, M., van den Bosch, F.C.,
Yang, X.H., More S., 2008, MNRAS, submitted

\bibitem[\protect\citeauthoryear{Longo}{2007}]{2007arXiv0707.3793L} Longo
M.~J., 2007, arXiv, 707, arXiv:0707.3793

\bibitem[\protect\citeauthoryear{Luki{\'c} et
al.}{2008}]{2008arXiv0803.3624L} Luki{\'c} Z., Reed D., Habib S.,
Heitmann K., 2008, arXiv, 803, arXiv:0803.3624

\bibitem[\protect\citeauthoryear{Mathis et al.}{2002}]{2002MNRAS.333..739M}
Mathis H., Lemson G., Springel V., Kauffmann G., White S.~D.~M.,
Eldar A., Dekel A., 2002, MNRAS, 333, 739

\bibitem[\protect\citeauthoryear{Mo
\& White}{1996}]{1996MNRAS.282..347M} Mo H.~J., White S.~D.~M.,
1996, MNRAS, 282, 347

\bibitem[\protect\citeauthoryear{Mo \& White}{2002}]{2002MNRAS.336..112M}
Mo H.~J., White S.~D.~M., 2002, MNRAS, 336, 112

\bibitem[\protect\citeauthoryear{Monaghan}{1992}]{1992ARA&A..30..543M} Monaghan J.~J., 1992, ARA\&A, 30, 543

\bibitem[\protect\citeauthoryear{Navarro, Frenk, \&
White}{1997}]{1997ApJ...490..493N} Navarro J.~F., Frenk C.~S.,
White S.~D.~M., 1997, ApJ, 490, 493

\bibitem[\protect\citeauthoryear{Peacock
\& Smith}{2000}]{2000MNRAS.318.1144P} Peacock J.~A., Smith R.~E.,
2000, MNRAS, 318, 1144

\bibitem[\protect\citeauthoryear{Pearce et al.}{2000}]{2000MNRAS.317.1029P}
Pearce F.~R., Thomas P.~A., Couchman H.~M.~P., Edge A.~C., 2000,
MNRAS, 317, 1029

\bibitem[\protect\citeauthoryear{Prada et al.}{2006}]{2006ApJ...645.1001P}
Prada F., Klypin A.~A., Simonneau E., Betancort-Rijo J., Patiri
S., Gottl{\"o}ber S., Sanchez-Conde M.~A., 2006, ApJ, 645, 1001

\bibitem[\protect\citeauthoryear{Press \&
Schechter}{1974}]{1974ApJ...187..425P} Press W.~H., Schechter P.,
1974, ApJ, 187, 425

\bibitem[\protect\citeauthoryear{Rauch}{1998}]{1998ARA&A..36..267R} Rauch
M., 1998, ARA\&A, 36, 267

\bibitem[\protect\citeauthoryear{Schmoldt et
al.}{1999}]{1999AJ....118.1146S} Schmoldt I.~M., et al., 1999, AJ,
118, 1146

\bibitem[\protect\citeauthoryear{Scranton}{2003}]{2003MNRAS.339..410S}
Scranton R., 2003, MNRAS, 339, 410

\bibitem[\protect\citeauthoryear{Sheth, Mo, \&
Tormen}{2001}]{2001MNRAS.323....1S} Sheth R.~K., Mo H.~J., Tormen
G., 2001, MNRAS, 323, 1

\bibitem[\protect\citeauthoryear{Somerville
\& Primack}{1999}]{1999MNRAS.310.1087S} Somerville R.~S., Primack
J.~R., 1999, MNRAS, 310, 1087

\bibitem[Spergel et al.(2007)]{2007ApJS..170..377S} Spergel, D.~N., et al.\
2007, \apjs, 170, 377

\bibitem[\protect\citeauthoryear{Springel}{2005}]{2005MNRAS.364.1105S}
Springel V., 2005, MNRAS, 364, 1105

\bibitem[\protect\citeauthoryear{Springel et
al.}{2005}]{2005Natur.435..629S} Springel V., et al., 2005, Natur,
435, 629

\bibitem[\protect\citeauthoryear{Stocke, Shull, \&
Penton}{2004}]{2004astro.ph..7352S} Stocke J.~T., Shull J.~M.,
Penton S.~V., 2004, astro, arXiv:astro-ph/0407352

\bibitem[\protect\citeauthoryear{Tinker et al.}{2008}]{2008arXiv0803.2706T}
Tinker J.~L, Kravtsov A.~V, Klypin A., Abazajian K., Warren M.~S,
Yepes G., Gottlober S., Holz D.~E, 2008, arXiv, 803,
arXiv:0803.2706

\bibitem[\protect\citeauthoryear{Tinker et al.}{2005}]{2005ApJ...631...41T}
Tinker J.~L., Weinberg D.~H., Zheng Z., Zehavi I., 2005, ApJ, 631,
41

\bibitem[\protect\citeauthoryear{Tripp
\& Bowen}{2005}]{2005pgqa.conf....5T} Tripp T.~M., Bowen D.~V.,
2005, pgqa.conf, 5

\bibitem[\protect\citeauthoryear{Vale
\& Ostriker}{2006}]{2006MNRAS.371.1173V} Vale A., Ostriker J.~P.,
2006, MNRAS, 371, 1173

\bibitem[\protect\citeauthoryear{van den Bosch}{2002}]{2002MNRAS.332..456V}
van den Bosch F.~C., 2002, MNRAS, 332, 456

\bibitem[\protect\citeauthoryear{van den Bosch, Yang,
\& Mo}{2003}]{2003MNRAS.340..771V} van den Bosch F.~C., Yang X.,
Mo H.~J., 2003, MNRAS, 340, 771

\bibitem[\protect\citeauthoryear{van den Bosch et
al.}{2007}]{2007MNRAS.376..841V} van den Bosch F.~C., et al.,
2007, MNRAS, 376, 841

\bibitem[\protect\citeauthoryear{Wang, Mo, \&
Jing}{2007}]{2007MNRAS.375..633W} Wang H.~Y., Mo H.~J., Jing
Y.~P., 2007, MNRAS, 375, 633

\bibitem[\protect\citeauthoryear{Wang et al.}{2007}]
{2007arXiv0711.4431W}
Wang, Y., Yang, X.H.,, Mo, H. J., van den Bosch, F.C.,
Weinmann, S. M., Chu, Y., 2007, arXiv0711.4431

\bibitem[\protect\citeauthoryear{Wechsler et
al.}{2006}]{2006ApJ...652...71W} Wechsler R.~H., Zentner A.~R.,
Bullock J.~S., Kravtsov A.~V., Allgood B., 2006, ApJ, 652, 71

\bibitem[\protect\citeauthoryear{Weinmann et
al.}{2006}]{2006MNRAS.366....2W} Weinmann S.~M., van den Bosch
F.~C., Yang X., Mo H.~J., 2006, MNRAS, 366, 2

\bibitem[\protect\citeauthoryear{White
\& Frenk}{1991}]{1991ApJ...379...52W} White S.~D.~M., Frenk C.~S.,
1991, ApJ, 379, 52

\bibitem[White \& Rees(1978)]{1978MNRAS.183..341W} White, S.~D.~M., \&
Rees, M.~J.\ 1978, \mnras, 183, 341

\bibitem[\protect\citeauthoryear{Yan, Madgwick,
\& White}{2003}]{2003ApJ...598..848Y} Yan R., Madgwick D.~S.,
White M., 2003, ApJ, 598, 848

\bibitem[\protect\citeauthoryear{Yang, Mo,
\& van den Bosch}{2003}]{2003MNRAS.339.1057Y} Yang X., Mo H.~J.,
van den Bosch F.~C., 2003, MNRAS, 339, 1057

\bibitem[\protect\citeauthoryear{Yang et al.}{2005}]{2005MNRAS.356.1293Y}
Yang X., Mo H.~J., van den Bosch F.~C., Jing Y.~P., 2005, MNRAS,
356, 1293

\bibitem[\protect\citeauthoryear{Yang et al.}{2007}]{2007ApJ...671..153Y}
Yang X., Mo H.~J., van den Bosch F.~C., Pasquali A., Li C., Barden
M., 2007, ApJ, 671, 153

\bibitem[\protect\citeauthoryear{York et al.}{2000}]{2000AJ....120.1579Y}
York D.~G., et al., 2000, AJ, 120, 1579

\bibitem[\protect\citeauthoryear{Zaroubi et
al.}{1995}]{1995ApJ...449..446Z} Zaroubi S., Hoffman Y., Fisher
K.~B., Lahav O., 1995, ApJ, 449, 446

\bibitem[\protect\citeauthoryear{Zheng et al.}{2005}]{2005ApJ...633..791Z}
Zheng Z., et al., 2005, ApJ, 633, 791

\end{thebibliography}
\end{document}